\documentclass[times,modern]{aastex63}

\usepackage[T1]{fontenc}
\usepackage{epsf,color,amsmath}

\usepackage{cancel}

\newcommand{\sfrac}[2]{\mathchoice%
  {\kern0em\raise.5ex\hbox{\the\scriptfont0 #1}\kern-.15em/
    \kern-.15em\lower.25ex\hbox{\the\scriptfont0 #2}}
  {\kern0em\raise.5ex\hbox{\the\scriptfont0 #1}\kern-.15em/
    \kern-.15em\lower.25ex\hbox{\the\scriptfont0 #2}}
  {\kern0em\raise.5ex\hbox{\the\scriptscriptfont0 #1}\kern-.2em/
    \kern-.15em\lower.25ex\hbox{\the\scriptscriptfont0 #2}} {#1\!/#2}}

\newcommand{\castro}{{\sf Castro}}
\newcommand{\maestro}{{\sf Maestro}}
\newcommand{\flash}{{\sf Flash}}
\newcommand{\chimera}{{\sf Chimera}}
\newcommand{\prompi}{{\sf PROMPI}}
\newcommand{\amrex}{{\sf AMReX}}

\newcommand{\isot}[2]{$^{#2}\mathrm{#1}$}
\newcommand{\isotm}[2]{{}^{#2}\mathrm{#1}}

\newcommand{\gcc}{\mathrm{g~cm^{-3} }}

\newcommand{\nablab}{{\mathbf{\nabla}}}
\newcommand{\Ub}{\mathbf{U}}
\newcommand{\gb}{\mathbf{g}}
\newcommand{\omegadot}{\dot{\omega}}
\newcommand{\Sdot}{\dot{S}}

\newcommand{\dedrd}{\left . \frac{\partial{}e}{\partial\rho}\right |_{T, X_k}}

\newcommand{\dedXd}{\left . \frac{\partial{}e}{\partial{}X_k} \right |_{\rho, T, X_{j,j\ne k}}}

\newcommand{\dedTd}{\left . \frac{\partial{}e}{\partial{}T} \right |_{\rho,X_k}}

\newcommand{\Ic}{\bm{\mathcal{I}}}
\newcommand{\kmax}{{K}}
\newcommand{\mmax}{{M}}
\newcommand{\kth}{k_\mathrm{th}}
\usepackage{bm}

\newcommand{\Uc}{{\,\bm{\mathcal{U}}}}
\newcommand{\fb}{\mathbf{f}}
\newcommand{\Fb}{\mathbf{F}}
\newcommand{\Sc}{\mathbf{S}}

\newcommand{\xv}{{(x)}}
\newcommand{\yv}{{(y)}}

\newcommand{\ex}{{\bf e}_x}
\newcommand{\ey}{{\bf e}_y}

\newcommand{\ijk}{{i,j}}

\newcommand{\qb}{{\bf q}}

\newcommand{\Shydro}{{{\bf H}}}
\newcommand{\Rb}{{\bf R}}
\newcommand{\Rbs}[1]{{\bf R} \left ( #1 \right )}

\newcommand{\Adv}[1]{{\left [\boldsymbol{\mathcal{A}} \left(#1\right)\right]}}

\newcommand{\Advs}[1]{\boldsymbol{\mathcal{A}} \left(#1\right)}
\newcommand{\Advsin}[1]{\boldsymbol{\mathcal{A}} (#1)}

\newcommand{\avg}[1]{{\left \langle #1 \right \rangle}}

\newcommand{\epsrho}{{\epsilon_{\mathrm{rel},\rho}}}
\newcommand{\epsspec}{{\epsilon_{\mathrm{rel},(\rho X)}}}
\newcommand{\epsener}{{\epsilon_{\mathrm{rel},(\rho e)}}}
\newcommand{\epsabs}{{\epsilon_{\mathrm{abs}}}}

\newcommand{\base}{{\mathrm{base}}}

\submitjournal{ApJ}

\begin{document}
\title{Improved Coupling of Hydrodynamics and Nuclear Reactions via Spectral Deferred Corrections}

\correspondingauthor{Michael Zingale}
\email{michael.zingale@stonybrook.edu}

\shorttitle{Coupling of hydrodynamics and Reactions}
\shortauthors{Zingale et al.}

\author[0000-0001-8401-030X]{M.~Zingale}
\affiliation{Dept.\ of Physics and Astronomy \\
             Stony Brook University \\
             Stony Brook, NY 11794-3800}

\author[0000-0003-0439-4556]{M.~P.~Katz}
\affiliation{NVIDIA Corporation}

\author[0000-0002-5749-334X]{J.~B.~Bell}
\affiliation{Center for Computational Sciences and Engineering \\
             Lawrence Berkeley National Laboratory \\
             Berkeley, CA  94720}

\author[0000-0002-0044-9778]{M.~L.~Minion}
\affiliation{Lawrence Berkeley National Laboratory \\
             Berkeley, CA  94720}

\author[0000-0003-1791-0265]{A.~J.~Nonaka}
\affiliation{Center for Computational Sciences and Engineering \\
             Lawrence Berkeley National Laboratory \\
             Berkeley, CA  94720}

\author[0000-0001-8092-1974]{W.~Zhang}
\affiliation{Center for Computational Sciences and Engineering \\
             Lawrence Berkeley National Laboratory \\
             Berkeley, CA  94720}

\begin{abstract}
Simulations in stellar astrophysics involve the coupling of
hydrodynamics and nuclear reactions under a wide variety of conditions, from
simmering convective flows to explosive nucleosynthesis.
Numerical techniques such as operator splitting (most notably Strang
splitting) are usually employed to couple the physical processes, but this
can affect the accuracy of the simulation, 
particularly when the burning is vigorous.  Furthermore,
Strang splitting does not have a straightforward extension to
higher-order integration in time.  We present a new temporal integration
strategy based on spectral deferred corrections and describe the second-
and fourth-order implementations in the open-source, finite-volume,
compressible hydrodynamics code \castro.
One notable advantage to these schemes is that they combine standard
low-order discretizations for individual physical processes in a way
that achieves an arbitrarily high order of accuracy.
We demonstrate the improved accuracy of the new methods
on several test problems of increasing complexity.
\end{abstract}


\section{Introduction}\label{Sec:Introduction}

Stellar astrophysical flows involve the delicate coupling of
hydrodynamics, reactions, and other physics (gravity, radiation,
magnetic fields, etc.).  Whether modeling convective or explosive
burning, reacting flows present temporal challenges to traditional
algorithms used in stellar / nuclear astrophysics.  Reaction network
ODE systems are often stiff\footnote{See \citealt{BYRNE19871} for some
definitions of stiffness.}, containing timescales
that are much smaller than hydrodynamic timescales.
For this reason, astrophysical hydrodynamics codes 
often employ operator splitting to couple the reactions and hydrodynamics, treating
the reactive portion of the evolution implicitly and the hydrodynamics
explicitly, and allowing each to take their preferred internal
timesteps.  Strang-splitting~\citep{strang:1968} is a widely used technique for
coupling in astrophysical systems (e.g., used in \castro,
\citealt{castro}, \maestro, \citealt{MAESTRO:Multilevel}, \flash,
\citealt{flash}, \chimera, \citealt{chimera}, \prompi,
\citealt{prompi}, and many others), but it can break down in regions
where energy is released faster than the hydrodynamics can respond.
Traditional Strang splitting is also limited to second-order accuracy in time, although higher-order variants are possible.  

As
astrophysical hydrodynamics codes push to higher-order spatial
accuracy (see, e.g., \citealt{apsara,athenapp,most:2019}), new time-integration
schemes are needed to realize the potential of high-order methods.
Here we look at alternate ways to couple hydrodynamics and reactions,
in particular, spectral deferred correction (SDC) methods.  We describe a
fully fourth-order method in space and time for
coupling hydrodynamics and reactions in the open source, 
finite-volume, compressible \castro~code.
One notable advantage of SDC schemes is that they combine
standard low-order discretizations for individual physical processes in a way that achieves an arbitrarily high order of accuracy.  
Here we evaluate these methods on a suite of test problems with the ultimate
goal of modeling thermonuclear flame
propagation in X-ray bursts, as we described in \citet{astronum:2018}.
In X-ray bursts, the range of lengthscales and tight hydrostatic equilibrium of
the atmosphere make models that capture the burning, flame scale, and
global scales of the neutron star challenging.  This problem is an
ideal candidate for higher-order methods.

The presentation in this paper is described as follows.
In \S~\ref{sec:model} we describe the model equations of interest.
In \S~\ref{sec:alg_overview} we present an overview of the Strang splitting methods and the second- and fourth-order SDC approaches.
In \S~\ref{sec:alg_details} we present the complete details of our SDC approach.
In \S~\ref{sec:tests} we demonstrate the accuracy of the new schemes on several different test problems, and in
\S~\ref{sec:summary} we discuss the strengths and weaknesses of the new scheme and future plans for extending the methods for more complex equations of interest.

\section{Model Equations}\label{sec:model}
The governing equations of interest in this work are based on
the fully compressible Euler equations, including
thermal diffusion.  Since our focus is on time-integration, we 
restrict the presentation to 1- and 2-d problems in this paper,
but the method is straightforward to extend to 3-d.
In conservation-law form, the 2-d system can be written
\begin{equation}
\Uc_t + [\Fb^\xv (\Uc)]_x + [\Fb^\yv (\Uc)]_y = \Sc(\Uc),
\end{equation}
where $\Uc = (\rho, (\rho X_k), (\rho \Ub), (\rho E), (\rho
e))^\intercal$ is the vector of conserved quantities, $\Fb^\xv$ and $\Fb^\yv$ are the
fluxes in the $x$- and $y$-directions, and $\Sc$ are source terms.  Here, $\rho$ is the mass density,
$\Ub$ is the velocity vector with components $u$ and $v$, $E$ is the
specific total energy, related to the specific internal energy, $e$,
as
\begin{equation}
E = e + |\Ub|^2/2
\end{equation}
and $X_k$ are the mass fractions of the reacting species, constrained
such that $\sum_k X_k = 1$.  The use of both $E$ and $e$ is
overdetermined; this is done for cases where calculating $e$ via $E -
|\Ub|^2/2$ yields an unreliable internal energy. One example is
roundoff error in regions of high Mach number flows
\citep{bryan:1995}.  The general stellar equation of state we use
here, with weak temperature-dependency in $e$ for
highly-degenerate gases also benefits from the dual-energy approach.
The use of both $\rho X$ and $\rho$ is also overdetermined, however
we integrate both to numerically ensure that we conserve total mass.

The fluxes are
\begin{equation}
\Fb^\xv (\Uc) = \left ( \begin{array}{c}
   \rho u \\ \rho X_k u \\ \rho u u + p \\ \rho v u \\ \rho u E + u p  \\ \rho u e
   \end{array} \right ) \, , \quad
\Fb^\yv (\Uc) = \left ( \begin{array}{c}
   \rho v \\ \rho X_k v \\ \rho u v \\ \rho v v + p \\ \rho v E + v p \\ \rho v e
   \end{array} \right ) \, , \quad
\end{equation}
Here the pressure, $p$, enters, and is found via the equation of state,
\begin{equation}
p = p(\rho, X_k, e)
\end{equation}
Finally, the source terms for our system typically include gravity,
thermal diffusion, and reactive terms.  For computational efficiency,
it is advantageous to treat the reactive terms separately from the other hydrodynamics source terms, so we
split this into two components, $\Shydro$, the hydrodynamic source,
and $\Rb$, the reactive source.  We note that since the $(\rho e)$
equation is not conservative, we include the ``$pdV$'' work term, $p \nabla\cdot \Ub$, as a source
term for that component. 
\begin{equation}
\label{eq:sources}
\Sc(\Uc) = \left ( \begin{array}{c}
   0 \\ \rho \omegadot_k \\ \rho \gb \cdot \ex \\ \rho \gb \cdot \ey 
        \\ \rho \Ub \cdot \gb + \rho \Sdot + \nabla \cdot \kth \nabla T\\ -p \nabla \cdot \Ub + \rho \Sdot + \nabla \cdot \kth \nabla T
 \end{array} \right )  =
\underbrace{\left ( \begin{array}{c}
   0 \\ 0 \\ \rho \gb \cdot \ex \\ \rho \gb \cdot \ey 
        \\ \rho \Ub \cdot \gb + \nabla \cdot \kth \nabla T\\ -p \nabla \cdot \Ub + \nabla \cdot \kth \nabla T
 \end{array} \right )}_{\Shydro(\Uc)} +
\underbrace{\left ( \begin{array}{c}
   0 \\ \rho \omegadot_k \\ 0 \\ 0 
        \\ \rho \Sdot \\ \rho \Sdot
 \end{array} \right )}_{\Rbs{\Uc}}
\end{equation}
Here, the species are characterized by mass fractions, $X_k$, and change via their creation rates, $\omegadot_k$, and the
energy has a corresponding specific energy generation rate, $\Sdot$.  The gravitational
acceleration, $\gb$, is found either by solving the Poisson equation,
\begin{equation}
\nabla^2 \Phi = 4\pi G \rho
\end{equation}
for the potential, $\Phi$, and then $\gb = -\nablab \Phi$, or by
externally specifying $\gb$, with the Cartesian unit vectors denoted
$\ex$ and $\ey$.  For the present work, we consider only constant
gravity (appropriate for our target XRB problem).  Since the reactions
and diffusion term depend on temperature, to close the system of
equations our equation of state also needs to return temperature, $T$,
\begin{equation}
T = T(\rho, X_k, e)
\end{equation}
The thermal conductivity, $\kth$, is likewise a function of the thermodynamic state:
\begin{equation}
\kth = \kth(\rho, X_k, T)
\end{equation}
We note that the thermal diffusion term could instead be included in
the definition of the fluxes, since it is represented as the
divergence of the diffusive flux.  We use this later in the
construction of a fourth-order approximation to thermal diffusion.

We rewrite our  system as:
\begin{equation}
\Uc_t = \Advs{\Uc} + \Rbs{\Uc}
\end{equation}
where we define the advective term, $\Advs{\Uc}$ to include the
hydrodynamic source terms:
\begin{equation}
\Advs{\Uc} \equiv - [\Fb^\xv (\Uc)]_x - [\Fb^\yv (\Uc)]_y + \Shydro(\Uc)
\end{equation}
We denote the discretized advective source with a cell subscript, as $\Adv{\Uc}_\ijk$.

\section{Algorithmic Overview}\label{sec:alg_overview}

\subsection{Strang Split Method}

A Strang-split integration method \citep{strang:1968} is an operator
splitting technique that alternates reactions and hydrodynamics, with each
indirectly seeing the effects of the other.
No direct coupling is
provided, but by staggering the operations, second-order accuracy is
achieved.  The basic algorithm to advance the solution over a time step $\delta t$
proceeds as:

\begin{itemize}
\item {\em Integrate the reactive portion of the system through $\delta t/2$}:
  We solve
  \begin{equation}
    \frac{d \Uc_\ijk}{dt} = \Rbs{\Uc_\ijk}
  \end{equation}
  starting with $\Uc^n$, yielding the solution $\Uc^\star$.  This is
  actually an ODE system, so we can use any implicit ODE integration
  scheme for this.  This evolution only sees the effects of reactions,
  not advection.  The species portion of the system has the form:
  \begin{equation}
    \label{eq:strang_reacting_system}
    \frac{dX_k}{dt} = \omegadot_k(\rho, T, X_k)
  \end{equation}
Accurate rate evaluation
  implies  we need  accurate temperature and density approximations, which suggests we
  should evolve the energy equation together with the species
  \citep{muller:1986}.  We can cast this in terms of temperature,
  \begin{equation}
    \label{eq:strang_temp_system}
    \rho c_x \frac{dT}{dt} = \rho \Sdot
  \end{equation}
  where $c_x$ is the specific heat.  This form is missing the
  ``$pdV$'' work term  $p \nabla \cdot \Ub$, since we are splitting the advective portion
  out of the reactive system.  For a constant volume burn, $c_x = c_v$
  is usually taken, while for a constant pressure burn, $c_x = c_p$ is
  taken (see, e.g., \citealt{ABNZ:III}).
  Again, if we were not
  splitting, then this choice would not matter, since the correct form
  of the work term would also appear.  Both
  Eqs.~\ref{eq:strang_reacting_system} and \ref{eq:strang_temp_system}
  are also missing the advective terms as well, so the extent to which the
  flow can transport fuel in and out of a cell is not accounted for
  during the burn.

\item {\em Advance the advective part through $\delta t$}:
  We solve
  \begin{equation}
    \frac{\partial \Uc_\ijk}{\partial t} = \Adv{\Uc}_\ijk
  \end{equation}
  integrating through the full $\delta t$, starting with $\Uc^\star$
  to get the state $\Uc^{n+1,\star}$.  No reaction sources are
  explicitly included, but since the first step integrated the effects
  of reactions to $\delta t/2$, the state that we build the advection
  from, $\Uc^\star$, is already time-centered with the effects of
  reactions, allowing us to be second-order in time (see
  \citealt{strang:1968} for a more formal discussion).

\item {\em Integrate the second half of reactions, through $\delta t/2$}:
  We again solve
  \begin{equation}
    \frac{d \Uc_\ijk}{dt} = \Rbs{\Uc_\ijk}
  \end{equation}
  starting with $\Uc^{n+1,\star}$, yielding the
  final solution $\Uc^{n+1}$.

\end{itemize}

With Strang splitting, there is no mechanism for the total density to
evolve during reactions (since that evolves solely through the advective terms), so the
burning is done at constant density.  Additionally, for stellar
material, the specific heat can be a strong function of temperature,
so we should allow the specific heat to evolve with the other
variables, which means augmenting the evolution of our system with an
equation of state call to keep the specific heat consistent. This is
sometimes neglected.

In \citet{astronum:2018},
the authors examine the numerical representation
of the solution over a time step for 
flow behind a detonation, comparing Strang
splitting to a simplified-SDC method.  This comparison shows that the
Strang {\em react-advect-react} sequence has strong departures from
equilibrium over the course of a step as compared to 
the simplified-SDC method.  The latter approach leads
to a smooth representation of the solution over a time step
due to improved advection/reaction coupling.  The
consequence of Strang splitting is that the departure from equilibrium 
requires much more computational work in the reaction steps 
due to the stiff transients
present as the system returns to equilibrium.  For
astrophysical reacting flow computations, the burning and EOS
components can require more computational effort than the hydrodynamic
components, so temporal methods that can reduce the number of
right-hand side calls in the reaction steps by staying closer to
equilibrium are advantageous.

We make one final note: the ODE system in the react step is done with
a specified numerical tolerance.  Tightening the tolerance means
solving the reacting system more accurately, but since we are
neglecting the hydrodynamics, we are in essence solving the wrong
equations very accurately.  So simply increasing the tolerances does
not necessarily lead to a more accurate solution, nor does it help 
improve the coupling with hydrodynamics.

\subsection{Method-of-Lines Integration}

Instead of splitting, we could discretize the entire system in space and 
then use an ODE integrator to handle the time evolution, a technique called
method-of-lines integration.
This would give us an ordinary differential
equations system to integrate of the form:
\begin{equation}
\frac{d\Uc_\ijk}{dt} = \Adv{\Uc}_\ijk + \Rbs{\Uc_\ijk}
\end{equation}
In an explicit MOL approach,
both the advective and reactive terms are evaluated with the same
state in constructing the right-hand side.  The difficulty here arises
if the reaction sources are stiff.  We need to use the same timestep
for both the reactions and hydrodynamics, and if we want to treat the
system explicitly (as we would like for hydrodynamics), then we are
constrained to evolve the entire system at the restrictive
timestep dictated by the reactions.  This can be computationally
infeasible.  Implicit and semi-implicit MOL approaches
(e.g., IMEX/Runge-Kutta approaches)
suffer from other maladies including the need to solve expensive coupled nonlinear equations, difficulty in generalizing to very high orders of accuracy, and difficulty in adding additional physical processes \citep{dutt:2000,minion:2003}.  With these limitations in mind, here we explore the SDC approach.

\subsection{General SDC Algorithm}

Generally, SDC algorithms are a class of numerical methods that represent the solution as an integral in time and iteratively solve a series of correction equations designed to reduce both the integration and splitting error. The correction equations are typically formed using a low-order time-integration scheme (e.g., forward or backward Euler), but are applied iteratively to construct schemes of arbitrarily high accuracy.
In practice, the time step is divided into a series of sub steps separated by nodes, and the solution at these nodes is iteratively improved by utilizing high-order integral representations of the solution from the previous iteration as a source term in the temporal integration.

The original SDC approach was introduced by \cite{dutt:2000} for ODEs where the integration of the ODE, as well as the associated correction equations, is performed using forward or backward Euler discretizations.
\cite{minion:2003} introduced a semi-implicit version (SISDC) for ODEs with stiff and non-stiff processes.  The correction equations for the non-stiff terms are discretized explicitly, whereas the stiff term corrections are treated implicitly.
While these works describe the solution of ODEs, they can be applied to PDEs 
discretizing in space and applying the method of lines.
Others have adapted this approach for multiphysics PDE simulation in a variety of contexts (3 time scales with substepping; finite volume/finite difference; compressible/incompressible and/or low Mach number reacting flow) to very high orders of accuracy (up to eighth-order); see \cite{BLM:2003,Almgren:2013,Emmett:2014,Pazner:2016,Emmett:2018}.
Here we leverage these ideas to develop a finite-volume, fourth-order spatio-temporal integrator for compressible astrophysical flow that couples reactions and hydrodynamics in a semi-implicit manner.

The basic idea of SDC is to write the solution of a system of ODEs
\begin{equation}
\frac{d\Uc}{dt} = \fb(t,\Uc(t)), \quad t\in[t^n, t^n + \delta t], \quad \Uc(t^n) \equiv \Uc^n,
\end{equation}
in the equivalent Picard  integral form,
\begin{equation}
\Uc(t) = \Uc^n + \int_{t^n}^t\fb(\Uc) d\tau,
\end{equation}
where we suppress explicit dependence of $\fb$ and $\Uc$ on $t$ for notational simplicity.  Given an approximation $\Uc^{(k)}(t)$ to $\Uc(t)$, the SDC correction equation is constructed by discretizing
\begin{equation}
\Uc^{(k+1)}(t) = \Uc^n + \int_{t^n}^t\left[\fb(\Uc^{(k+1)}) - \fb(\Uc^{(k)})\right] d\tau + \int_{t^n}^t\fb(\Uc^{(k)})d\tau, \label{eq:SDC_correction}
\end{equation}
where a low-order discretization (e.g., forward or backward Euler) is used for the first integral and a higher-order quadrature is used to evaluate the second integral.  Each iteration improves the overall order of accuracy of the approximation by one per iteration, up to the order of accuracy of the underlying quadrature rule used to evaluate the second integral.

\replaced{Numerically, for a given time step, we define $t^{n+1} = t^n
  + \delta t$, and divide the time step into $\mmax$ subintervals
  using $\mmax+1$ temporal nodes, so that $t^n \equiv t^0 < t^1 <
  \cdots < t^{\mmax} \equiv t^{n+1}$ and $\delta t_m = t^{m+1} - t^m$.}{For a given time step, we begin with a state at time $t^n$
  and seek to updated it through $\delta t$ to time $t^{n+1}$.  For the SDC
  method, we divide this time interval into $\mmax$ subintervals
  using $\mmax+1$ temporal nodes, $t^0 < t^1 <
  \cdots < t^{\mmax}$.  For the cases presented here, one 
  time node is at the beginning of the full time step ($t^0 \equiv t^n$) and one of at 
  the end of the full time step $t^M \equiv t^{n+1}$.}
  For a fourth-order approach, we can use 3-point Gauss-Lobatto
  quadrature with nodes located at the beginning, midpoint, and end of
  the time step. Using the notation $\Uc^{m,(k)}$ to denote the
  $k^{\rm th}$ iterate of the solution at node $m$, we can generalize
  (\ref{eq:SDC_correction}) to update the solution at a particular
  node
\begin{equation}
\Uc^{m+1,(k+1)} = \Uc^{m,(k+1)} + \int_{t^m}^{t^{m+1}}\left[\fb(\Uc^{(k+1)}) - \fb(\Uc^{(k)})\right] d\tau + \int_{t^m}^{t^{m+1}}\fb(\Uc^{(k)})d\tau
\end{equation}
Thus, a pure forward-Euler discretization of the first integral results in the update,
\begin{equation}
\Uc^{m+1,(k+1)} = \Uc^{m,(k+1)} + \delta t_m\left[\fb(\Uc^{m,(k+1)}) - \fb(\Uc^{m,(k)})\right] + \int_{t^m}^{t^{m+1}}\fb(\Uc^{(k)})d\tau
\end{equation}
Note that overall this is an explicit computation of $\Uc^{m+1,(k+1)}$ since all the terms on the right-hand side are available via explicit computation \added{($\Uc^{m,k+1}$ is at the current iteration but from the previous time node while $f(\Uc^{k})$ in the integral uses only the information from the previous iteration)}.

Our model contains advection and reaction terms,
\begin{equation} 
\Uc^{n+1} = \Uc^n + \int (\Advs{\Uc} + \Rbs{\Uc} ) dt
\end{equation}
Stiff terms (e.g. reactions) can be treated implicitly while non-stiff
terms (e.g. the hydrodynamics) can be treated explicitly.  Following \cite{minion:2003},
the correction equation contains an explicit and implicit part:
\begin{eqnarray}
\Uc^{m+1,(k+1)} = \Uc^{m,(k+1)} 
&+& \delta t_m\left[\Advs{\Uc^{m,(k+1)}} - \Advs{\Uc^{m,(k)}}\right] +\nonumber\\
&+& \delta t_m\left[\Rbs{\Uc^{m+1,(k+1)}} - \Rbs{\Uc^{m+1,(k)}}\right]\nonumber\\
&+& \int_{t^m}^{t^{m+1}}\Advs{\Uc^{(k)}} + \Rbs{\Uc^{(k)}}d\tau.
\end{eqnarray}
Now we note that overall this is an implicit equation for reactions solving for $\Uc^{m+1,(k+1)}$; all terms on the right-hand side except for $\Rb(\Uc^{m+1,(k+1)})$ are available via explicit computation.  Thus, the update at each node amounts to solving an implicit reaction equation with explicitly computed source terms from advection, as well as previously computed reaction terms.

\section{Algorithmic Details}\label{sec:alg_details}

Here we describe the fourth-order algorithm in full detail.
For proper construction of fourth-order methods, we need to distinguish between
cell-average values and cell-center values (i.e., point values) in the finite-volume
framework.  We use angled braces, e.g.~$\avg{\Uc}_\ijk$, to denote a cell average and $\Uc_\ijk$ to denote
a cell-center value.  For expressions involving the time-update for a single zone, we will drop
the spatial subscripts, $\ijk$.
In the SDC equations, we denote the state
with two time superscripts, $\avg{\Uc}^{m,(k)}$, where $m$ represents
the quadrature point in the time-discretization and $k$ represents the
iteration number.  A general SDC update for our state $\avg{\Uc}$
takes the form:
\begin{itemize}
\item For all SDC iterations $k \in [0,\kmax-1]$, we initialize the state, advective update, and reaction source for $m=0$ equal to the state at $t^n$ (i.e., the state at the $m=0$ node doesn't change with iteration and is equal to the state at $t^n)$;
  \begin{align}
    \avg{\Uc}^{0,(k)} &= \avg{\Uc}^n \\
    \avg{\Advs{\Uc}}^{0,(k)} &= \avg{\Advs{\Uc}}^n \\
    \avg{\Rbs{\Uc}}^{0,(k)} &= \avg{\Rbs{\Uc}}^n 
  \end{align}
\item We need values of $\Advs{\Uc}$ and $\Rb(\Uc)$ at all time nodes
  to do the integral over the iteratively-lagged state in the first
  iteration.  For all temporal nodes $m \in [1,\mmax]$, we initialize the
  state, advective update, and reaction source for $k=0$ equal to the
  state at $t^n$ (i.e., copy the $t^n$ state into each temporal node
  to initialize the time step).
  \begin{align}
    \avg{\Uc}^{m,(0)} &= \avg{\Uc}^n \\
    \avg{\Advs{\Uc}}^{m,(0)} &= \avg{\Advs{\Uc}}^n \\
    \avg{\Rbs{\Uc}}^{m,(0)} &= \avg{\Rbs{\Uc}}^n 
  \end{align}

\item Loop from $k = 0, \ldots, \kmax-1$

  This is the main iteration loop, and each pass  results in an
  improved value of $\avg{\Uc}$ at each of the time nodes.

\begin{itemize}

  \item Loop over time nodes, from $m = 0, \ldots, \mmax-1$

  \begin{itemize}

    \item Compute $\avg{\Advs{\Uc}}^{m,(k)}$ from $\Uc^{m,(k)}$

    \item Solve:
    \begin{align}
      \label{eq:sdc:general}
      \avg{\Uc}^{{m+1},(k+1)} = \avg{\Uc}^{m,(k+1)}
            &+ \delta t_m \left [ \avg{\Advs{\Uc}}^{m,(k+1)} - \avg{\Advs{\Uc}}^{m,(k)} \right ] \nonumber \\
            &+ \delta t_m \left [ \avg{\Rbs{\Uc}}^{{m+1},(k+1)} - \avg{\Rbs{\Uc}}^{{m+1},(k)} \right ] \nonumber \\
            &+ \Ic_m^{m+1}\left  ( \avg{\Advs{\Uc}}^{(k)} + \avg{\Rbs{\Uc}}^{(k)}\right ) 
    \end{align}
    where $\delta t_m = t^{m+1} - t^m$.  \added{Here, $\avg{\Rbs{\Uc}}$ is
    computed by evaluating the reactive source term,
    Eq.~\ref{eq:sources}, with the state $\Uc$ at the specified time
    node and iteration (details for constructing the spatial average are discussed in section~\ref{sec:fourth_burn}).}   The last term is an integral:
   \begin{align}
     \Ic_m^{m+1}\left ( \avg{\Advs{\Uc}}^{(k)} + \avg{\Rbs{\Uc}}^{(k)}\right ) & \approx \nonumber \\
        \int_{t^m}^{t^{m+1}} dt & \left (\avg{\Advs{\Uc}}^{(k)} + \avg{\Rbs{\Uc}}^{(k)}\right)
   \end{align}
   which we evaluate using an appropriate numerical quadrature rule
   with our $\mmax+1$ integration points, using the right hand side values from the
   previous iteration.
  \end{itemize}

\end{itemize}
\end{itemize}

We note that Eq.~\ref{eq:sdc:general} is an implicit nonlinear
equation for $\avg{\Uc}^{m+1,(k+1)}$.  There are four terms on the
right-hand side.  The second term is the increment in advection from
one SDC iteration to the next, written as a forward Euler (explicit)
update to the solution from one time quadrature point to the next.
The third term (the increment in the reaction term) appears as a
backwards Euler (implicit) update to the solution.  Note the reaction
terms here are the instantaneous rates---there is no ODE integration.
If the SDC iterations converge, then the increment in advection
and reactions (the second and third terms) go to zero, and the SDC
substeps tend toward:
\begin{equation}
      \avg{\Uc}^{{m+1},(k+1)} \approx \avg{\Uc}^{m,(k+1)}  +
\int_{t^m}^{t^{m+1}} \mathrm{d}t \left (\avg{\Advs{\Uc}}^{(k)} + \avg{\Rbs{\Uc}}^{(k)}\right).
\end{equation}
Hence the SDC solution converges to that of a
fully implicit Gauss-Runge-Kutta (also referred to as collocation)
method \citep{HuangJiaMinion06}.
This implies that the integrals that span the full timestep 
ultimately determine the accuracy of the method, and we pick the
number of quadrature nodes $M+1$,  to yield the desired time
accuracy over the timestep.  Lobatto methods have formal order $2M$, hence
we use 2 nodes for second-order in time, and 3 nodes for fourth-order in time.
We also see that during the SDC substeps for
each process (advection and reacting), terms 
coupling other processes are included, giving us the strong
coupling we desire.



Finding the new value of $\avg{\Uc}$ requires a nonlinear solve of
\begin{equation}
  \label{eq:fourthorderupdate}
    \avg{\Uc}^{m+1,(k+1)} - \delta t_m \avg{\Rbs{\Uc}}^{m+1,(k+1)} = \avg{\Uc}^{m,(k+1)} + \delta t_m \avg{\bf C}
\end{equation}
where the right-hand side is constructed only from known states, and we
define $\avg{\bf C}$ for convenience as:
\begin{align}
\avg{\bf C} &= \left [ \avg{\Advs{\Uc}}^{m,(k+1)} - \avg{\Advs{\Uc}}^{m,(k)} \right ] 
               -  \avg{\Rbs{\Uc}}^{{m+1},(k)}  \nonumber \\
            &+ \frac{1}{\delta t_m} \Ic_m^{m+1}\left  ( \avg{\Advs{\Uc}}^{(k)} + \avg{\Rbs{\Uc}}^{(k)}\right )
\end{align}
and note that it represents an average over the cell.

\subsection{Second-order Algorithm}

For second-order, we need only 2 quadrature points (M = 1), $m = 0$
corresponding to the old time solution, and $m = 1$ corresponding to
the new time solution.  In this case, the second term in
Eq.~\ref{eq:sdc:general} (the increment in advection terms from one
iteration to the next) cancels, since regardless of the iteration,
$k$, the solution at the old time ($m = 0$) is the same.  We can
approximate our integral using the trapezoid rule, which is second-order
accurate. We drop the $m$ superscript in what follows and simply denote
the solution as either at time-level $n$ or $n+1$.  Finally, to second-order
accuracy in space, we can take the cell-averages to be
cell-centers, and write $\avg{\Uc}_\ijk = \Uc_\ijk$.  The overall
integration is then:

\begin{itemize}

\item Loop from $k = 0, \ldots, \kmax-1$

  \begin{itemize}
  \item Solve:
    \begin{align}
    \label{eq:sdc:general_2nd}
      \Uc^{n+1,(k+1)} = \Uc^n &+ \delta t \left [ \Rbs{\Uc^{n+1,(k+1)}} - \Rbs{\Uc^{n+1,(k)}} \right ] \nonumber \\
                            &+ \frac{\delta t}{2} \left [ \Advs{\Uc^n} + \Advs{\Uc^{n+1,(k)}} +
                          \Rbs{\Uc^n} + \Rbs{\Uc^{n+1,(k)}} \right ]
    \end{align}

   We note that this is an implicit nonlinear equation for $\Uc^{n+1,(k+1)}$.

  \item Using this $\Uc^{n+1,(k+1)}$, compute $\Advsin{\Uc^{n+1,(k+1)}}$ and
    $\Rb(\Uc^{n+1,(k+1)})$ for use in the next iteration.
  \end{itemize}

\end{itemize}
Choosing $\kmax = 2$ is sufficient for second-order accuracy.  

To construct the advective term, $\Advs{\Uc}$, we use piecewise linear
reconstruction with the slope limiter from \citet{colella:1985}.  The
reconstructed slopes give us the edge states for each interface and we
solve the Riemann problem to get the fluxes through the interface
state.

To solve the update to second-order, we can ignore the difference
between cell centers and cell-averages, and simply solve an implicit 
equation of the form:
\begin{equation}
\label{eq:second_order_update}
  {\bf Z}(\Uc^{n+1}) = \Uc^{n+1} - \delta t_m \Rbs{\Uc^{n+1}} - \Uc^n  - \delta t_m {\bf C} = 0
\end{equation}
(we drop the iteration index, $k$, on the unknown here, and use the
old and new time levels, $n$ and $n+1$ since there are no intermediate
time nodes).  This allows the update of a cell to be independent of
other cells.

\subsubsection{Solving the nonlinear system}

We use a simple Newton iteration to solve 
Eq.~\ref{eq:second_order_update}.  As we discuss below, 
an accurate initial guess may be required for the iteration to converge, especially for complex or very
stiff networks.  Given an initial guess for the solution, $\Uc_0$, we seek a
correction, $\delta \Uc$, as
\begin{equation}
  \label{eq:linearized_system}
        {\bf Z}(\Uc_0 + \delta \Uc) = {\bf Z}(\Uc_0) + {\bf J} \delta \Uc + \ldots \approx 0
\end{equation}
Here, {\bf J} is the Jacobian, $\partial {\bf Z}/\partial \Uc$.  For
the reactions, a more natural representation to work in is ${\bf w} = (\rho,
X_k, \Ub, T)^\intercal$, as reaction networks typically provide a
Jacobian in these terms.  Our full Jacobian is then:
\begin{equation}
  {\bf J} \equiv \frac{\partial {\bf Z}}{\partial \Uc} = {\bf I} - \delta t_m \frac{\partial \Rb}{\partial \Uc}
  = {\bf I} - \delta t_m \frac{\partial \Rb}{\partial {\bf w}} \frac{\partial {\bf w}}{\partial \Uc}
\end{equation}
where we can compute $\partial {\bf w}/\partial \Uc$ as $(\partial
\Uc/\partial {\bf w})^{-1}$, and the reaction network provides
$\partial \Rb/\partial {\bf w}$.

Solving Eq.~\ref{eq:linearized_system} requires solving the linear system
${\bf J} \delta \Uc = -{\bf Z}(\Uc_0)$.
In practice, we only need to do the implicit solve for $\Uc' = (\rho,
\rho X_k, \rho e)^\intercal$ with ${\bf w}' = (\rho, X_k,
T)^\intercal$, reducing the size of the Jacobian, and we can update
the momentum, $\rho \Ub$, explicitly.  We always compute $\partial
{\bf w}/\partial \Uc$ analytically, but $\partial \Rb/\partial {\bf
  w}$ is  computed either numerically, via differencing, or
analytically, depending on the reaction network.
Appendix~\ref{sec:app:jac} gives the form of these matrices.
We solve iteratively, applying the correction $\delta \Uc$
to our guess $\Uc_0$ until $\delta \Uc$ satisfies
\begin{equation}
  \left \{ \frac{1}{N+2} \left [
    \left | \delta \Uc(\rho) \,w_\rho \right |^2 + 
    \left | \delta \Uc(\rho e) \,w_{(\rho e)} \right |^2 + 
    \sum_{k=1}^N \left | \delta \Uc(\rho X_k) \,w_{(\rho X_k)} \right |^2 \right ] \right \}^{1/2} < 1,
\end{equation}
We specify separate relative tolerances for the density, species, and
energy, $\epsrho$, $\epsspec$, and $\epsener$ respectively, as well as
an overall absolute tolerance, $\epsabs$, and $N$ is the number of
species in the network.  We use a tolerance comparable to what we
would use when integrating a reaction network directly.  For each of
the state components of $\Uc$ we define a weight the form:
\begin{align}
w_{\rho} &= \left [{\epsrho | \Uc(\rho) | + \epsabs} \right ]^{-1} \enskip , \\
w_{(\rho e)} &= \left [ {\epsener | \Uc(\rho e) | + \epsabs} \right ]^{-1} \enskip , \\
w_{(\rho X_k)} &= \left [ {\epsspec | \Uc(\rho X_k) | + \epsabs |\Uc(\rho)|} \right ]^{-1} \enskip .
\end{align}
This is inspired by the convergence measure used by VODE~\citep{vode}.
Note that for the mass fractions, we scale the absolute tolerance,
$\epsabs$, by the density, since we evolve partial densities.  The
absolute tolerance is critical to ensuring that our solve does not
stall due to trace species in the network.  For the other quantities,
the only purpose of $\epsabs$ is to prevent a division by zero.
While not strictly necessary, we re-evaluate the reaction source with
the solution after the Newton solve,
$\Uc^{m+1,(k+1),\star}$, to get the final update:
\begin{equation}
  \Uc^{m+1,(k+1)} = \Uc^{m,(k+1)} + \delta t_m \Rbs{\Uc^{m+1,(k+1),\star}} + \delta t_m {\bf C}
\end{equation}

\subsubsection{Initial guess for the nonlinear system}

A simple initial guess can be made by extrapolating in time for the
first SDC iteration and using the result from the previous iteration for
the new time node during later iterations:
\begin{equation}
  \Uc_0 = \begin{cases}
    \Uc^{m,(0)} + \delta t_m \left [ \Advs{\Uc^{m,(0)}} + \Rbs{\Uc^{m,(0)}}\right ] & \mbox{if}~k = 0 \\
    \Uc^{m+1,(k)} & \mbox{if}~k > 0
  \end{cases}
\end{equation}

If a single Newton step does not converge, we
subdivide the interval $[0, \delta t_m]$ into a number of substeps
(starting with 2, and continuing to double the number of substeps until convergence or we reach a 
specific limit---we use 64)
and do a backwards difference update for each substep, each of
which would take the form of the simple Newton iteration described
above.

For very stiff problems, Newton iterations may fail to converge.
\cite{Emmett:2018} casts Equation~\ref{eq:second_order_update} as an ODE, 
since it is essentially a backwards difference update for $\Uc$,
writing it as:
\begin{equation}
  \frac{d\Uc}{dt} \approx \frac{\Uc^{m+1} - \Uc^m}{\delta t_m} = \Rbs{\Uc} + {\bf C}
\end{equation}
and uses a stiff ODE solver (like VODE, \citealt{vode}) to integrate
it from one time node to the next for the first iteration.  The
initial condition for the integration is $\Uc^{m,(k)}$.  A stiff ODE
solver requires the Jacobian corresponding to the right-hand side of
the system, which is simply
\begin{equation}
  {\bf J} = \frac{\partial \Rb}{\partial \Uc} = \frac{\partial \Rb}{\partial {\bf w}} \frac{\partial {\bf w}}{\partial \Uc}
\end{equation}
We would still provide VODE with the same set of tolerances defined
above, $\epsrho$, $\epsspec$, and $\epsener$, as well as an absolute
tolerance for the species, $\epsabs \rho^{m,(k)}$.  The solution from
VODE is then used in Eq.~\ref{eq:second_order_update}.  This method
converges well but is not needed for the problems presented here.  We
will explore it further in subsequent studies.

\subsection{Fourth-order Algorithm}

To construct a fourth-order temporal discretization, we use three-point Gauss-Lobatto quadrature in time that introduces a quadrature node at the midpoint.  We consider
the time at $t^m$, where $m = 0, 1, 2$, and $m = 0$ corresponds to the
start time, $t^n$, $m = 1$ corresponds to the midpoint, $t^{n+1/2}$,
and $m = 2$ corresponds to $t^{n+1}$, the new-time solution.  We also
take $\kmax = 4$.  For the integral, $\mathcal{I}_m^{m+1}$,
we use Simpson's rule by deriving the weights for a parabola passing
through the points at $t^0$, $t^1$, and $t^2$, and then
constructing the integral over the desired sub-interval.  This derivation is presented
in the supplemental Jupyter notebook, and results in:
\begin{equation}
\mathcal{I}_{0}^{1} = \int_{t^0}^{t^1} \phi(t) dt = \frac{\delta t}{24} (5 \phi_0 + 8\phi_1 - \phi_2) \enskip ,
\end{equation}
and
\begin{equation}
\mathcal{I}_{1}^{2} = \int_{t^1}^{t^2} \phi(t) dt = \frac{\delta t}{24} (-\phi_0 + 8\phi_1 +5 \phi_2) \enskip ,
\end{equation}
where $\phi_m\equiv\phi(t^m)$.

\subsubsection{Spatial Discretization}

To construct a fourth-order accurate finite-volume discretization,
the difference between cell-centers and cell-averages is important.  We can find a relation between the two by
starting with the definition of a cell average of a function $f(x)$, which in 1-d is
\begin{equation}
\langle f\rangle_i = \frac{1}{h} \int_{x_i - h/2}^{x_i + h/2} f(x) dx
\end{equation}
where $h$ is the width of the cell.  Taylor expanding $f(x)$ about the cell-center, $x_i$, as:
\begin{equation}
f(x) = \sum_{n=0}^\infty \frac{f^{(n)}(x_i)}{n!} (x - x_i)^n
\end{equation}
The odd terms integrate to zero, and to fourth order we have:
\begin{equation}
\langle f\rangle_i = f(x_i) + \frac{h^2}{24} \left .\frac{d^2{}f}{dx^2} \right |_{x_i} + \mathcal{O}(h^4)
\end{equation}
In the multi-dimensional extension, the second derivative becomes a
Laplacian.  A similar construction can be used to convert a
face-centered quantity to a face-averaged quantity.  These ideas were
used in \cite{mccorquodalecolella} to construct a fourth-order
accurate finite volume method for compressible hydrodynamics.  We briefly
summarize their reconstruction here:
\begin{enumerate}
  \item Compute a cell-average primitive variable state $\avg{\qb}_\ijk$
    from the conserved cell-average state, $\avg{\Uc}_\ijk$.

  \item Reconstruct the cell-average state $\avg{\qb}_\ijk$ to edges to
    define a face-average interface state, $\avg{\qb}_{i+1/2,j}$.  Note
    that if limiting is done in the reconstruction, then a Riemann
    problem is solved here to find the unique interface state.

  \item Compute a face-centered interface state, $\qb_{i+1/2,j}$, from
    the face-averaged interface state, $\avg{\qb}_{i+1/2,j}$,
    \begin{equation}
      \qb_{i+1/2,j} = \avg{\qb}_{i+1/2,j} - \frac{h^2}{24} \Delta^{(2,f)} \avg{\qb}_{i+1/2,j}
    \end{equation}
    where $\Delta^{(2,f)}$ is the Laplacian only in the transverse
    direction (within the plane of the face).

  \item Evaluate the fluxes using both the face-average state, $\Fb(\avg{\qb}_{i+1/2,j})$,
    and the face-center state, $\Fb(\qb_{i+1/2,j})$ and compute the final face-average
    flux, $\avg{\Fb}_{i+1/2,j}$:
    \begin{equation}
      \label{eq:flux_avg}
      \avg{\Fb}_{i+1/2,j} = \Fb(\qb_{i+1/2,j}) + \frac{h^2}{24} \Delta^{(2,f)} \Fb(\avg{\qb}_{i+1/2,j})
    \end{equation}

\end{enumerate}

Here we adopt the notation of \citet{mccorquodalecolella}, and use
$\Delta^{(2)}$ to mean a second-order accurate discrete approximation
to the Laplacian operator.
The advection term is then
\begin{equation}
\left [ \Advs{\Uc} \right]_\ijk = - \frac{\avg{\Fb}^\xv_{i+i/2,j} - \avg{\Fb}^\xv_{i-1/2,j}}{h}
             - \frac{\avg{\Fb}^\yv_{i,j+1/2} - \avg{\Fb}^\yv_{i,j-1/2}}{h} + \avg{\Shydro(\Uc)}_\ijk.
\end{equation}
We implement the
spatial reconstruction as described there, including flattening and
artificial viscosity.  For the Riemann problem, we use the two-shock
solver that is the default in \castro\ (see \citealt{castro}).  Instead of 
the Runge-Kutta  method used in \citet{mccorquodalecolella}, we instead use
the SDC algorithm described above.

For physical boundaries, we follow the prescription in
\citet{mccorquodalecolella} to use one-sided stencils for the initial
reconstruction of the interface states to avoid needing information
from outside the domain.  We enforce reflecting boundary conditions on
the interface states at the boundary by reflecting the interior edge
state across the domain boundary (changing the sign of the normal
velocity).  This  forces the Riemann problem to give zero flux
through the interface.  For the Laplacian used in the transformation
between cell-centers and averages and face-centers and averages, we
differ slightly from their prescription, opting instead to use
one-sided second-order accurate differences for any direction in the
Laplacian that would reach across the domain boundary.

There are a few changes to support a general EOS and reactions to fourth-order
in space, which we summarize here:
\begin{itemize}
  \item {\em Treatment of $\Gamma_1$}:

    The Riemann problem uses the speed of sound, $c$, computed as
    \begin{equation}
      c = \left (\frac{\Gamma_1 p}{\rho}\right )^{1/2}
    \end{equation}
    for the left and right state as part of its solution.  This means that
    we need an interface value of $\Gamma_1 = d\log p /d\log \rho |_s$.
    The fourth-order solver does a single Riemann solve to get the state
    on the interfaces, starting with the interface states constructed
    from the cell-averages---we think of this as a face-average.  We
    perform the same reconstruction on $\Gamma_1$ and treat it as
    part of the interface state for this Riemann solve.

  \item {\em Hydrodynamic source terms}: For general (non-reacting)
    source terms, we  first convert the state variables to
    cell-centers,
    \begin{equation}
      \Uc_\ijk = \avg{\Uc}_\ijk - \frac{h^2}{24} \Delta^{(2)} \avg{\Uc}_\ijk
    \end{equation}
    We then evaluate the source terms point-wise using this:
    \begin{equation}
      \Shydro_\ijk = \Shydro(\Uc_\ijk)
    \end{equation}
    and finally convert the sources to cell-averages:
    \begin{equation}
      \avg{\Shydro}_\ijk = \Shydro_\ijk + \frac{h^2}{24} \Delta^{(2)} \Shydro_\ijk.
    \end{equation}
    We only need $\avg{\Shydro}_\ijk$ in for the interior cells.

    The first explicit source we consider here comes from the
    dual-energy formulation where we carry an internal energy
    evolution equation.  The internal energy evolution follows:
    \begin{equation}
      \frac{\partial (\rho e)}{\partial t} + \nabla \cdot (\rho e \Ub) + p \nabla \cdot \Ub = \nabla \cdot \kth \nabla T
    \end{equation}
    The term, $p \nabla \cdot \Ub$, is not in conservative form,
    so we cannot construct it to fourth order following the same
    procedure as we do with the fluxes.
    Instead, we add this term to
    the hydrodynamic sources and treat it as described above.  The constant gravity source is treated the same way.

    We note that for the SDC integration, we do not include the
    thermal diffusion term in the source terms $\Shydro$, but instead
    add them to the flux directly, as described below.

  \item {\em Deriving temperature / resetting $e$ with a real EOS}: Astrophysical
    equations of state are often posed with density and temperature as
    inputs, so the process of obtaining the temperature given density
    and specific internal energy requires an inversion, usually using
    Newton-Raphson iteration.  This results in a state that is
    thermodynamically consistent to some tolerance (we use a tolerance
    of $10^{-8}$).  It is important to leave the input $e$ unchanged
    after the EOS call, even though it may not be consistent with the
    EOS for the $T$ obtained via the Newton-Raphson iterations because
    of the tolerance used.  We consider the internal energy from its separate
    evolution here as well, and whether it should reset the internal energy
    derived from the total energy.  

    The overall procedure we use is:
    \begin{enumerate}
      \item Convert the cell-average conserved state to cell centers as:
        \begin{equation}
          \Uc_\ijk = \avg{\Uc}_\ijk - \frac{h^2}{24} \Delta^{(2)} \avg{\Uc}_\ijk
        \end{equation}

      \item Consider $e_\ijk$ as obtained from the separate internal
        energy evolution and if needed, reset the total energy to $E_\ijk
        = e_\ijk + |\Ub_\ijk|^2/2$, according to the procedure described in \citet{wdmergerI}.

      \item Compute the temperature from $\Uc_\ijk$.

      \item Convert back to cell-averages (we only need to do this for $T$ and $(\rho e)$) as:
        \begin{equation}
          \avg{\Uc}_\ijk = \Uc_\ijk + \frac{h^2}{24} \Delta^{(2)} \avg{\Uc}_\ijk
        \end{equation}
        Note: we use exactly the same Laplacian term here as
        in step 1 above---this is essential, since if the internal energy reset
        does nothing, this  leaves the energy unchanged to
        roundoff error.  If we instead constructed the Laplacian as
        $\Delta^{(2)} \Uc_\ijk$, it would not cancel out the previous
        Laplacian term, leading to an error at truncation level that
        builds up over the simulation.

    \end{enumerate}

    \item {\em Thermal diffusion}.  We include the diffusive flux in
      the energy flux, rewriting them as:
      \begin{align}
         \Fb^\xv (\rho E) &= \rho u E + u p - \kth \frac{\partial T}{\partial x} \\
         \Fb^\xv (\rho e) &= \rho u e - \kth \frac{\partial T}{\partial x} 
      \end{align}
      and similarly for the $(\rho E)$ and $(\rho e)$ components of
      $\Fb^\yv$.  Consider the x-direction.  We need to
      add this diffusive term to both $\Fb^\xv(\qb_{i+1/2,j})$ and
      $\Fb^\xv(\avg{\qb}_{i+1/2,j})$ before they are combined to make
      the final face-average flux, $\avg{\Fb^\xv}_{i+1/2,j}$ as in
      Eq.~\ref{eq:flux_avg}.  We compute these terms using the following discretizations:
      \begin{enumerate}
        \item {\em Flux from face-center state}.  Since we are computing the flux from
          the face-center quantity, we want to compute $\partial T/\partial x$ using
          cell-center values of the temperature, $T_\ijk$.  A fourth-order accurate 
          discretization of the first derivative, evaluated on the interface is
          \begin{equation}
            \left . \frac{\partial T}{\partial x} \right |_{i+1/2,j} =
            \frac{-T_{i+2,j} + 27 T_{i+1,j} - 27 T_{i,j} + T_{i-1,j}}{24 h}
          \end{equation}
          The thermal conductivity is simply evaluated from the
          face-center primitive variable state resulting from the
          Riemann solver, giving the diffusive flux:
          \begin{equation}
            \Fb^\xv_{\mathrm{diffusive}}((\rho e)_{i+1/2,j}) = - \kth(\qb_{i+1/2,j}) \left . \frac{\partial T}{\partial x} \right |_{i+1/2,j}
          \end{equation}

        \item {\em Flux from face-average state}.  We need to compute
          the face-average temperature gradient from the cell-average
          temperatures.  This is done by constructing a cubic
          conservative interpolant, differentiating it, and evaluating
          it at the desired interface, giving:
          \begin{equation}
            \left . \frac{\partial \avg{T}}{\partial x} \right |_{i+1/2,j} =
            \frac{-\avg{T}_{i+2,j} + 15 \avg{T}_{i+1,j} - 15 \avg{T}_{i,j} + \avg{T}_{i-1,j}}{12 h}
          \end{equation}
          We note this same expression appears in \citet{KADIOGLU20082012} and other sources.
          We use the face-average primitive variable state to evaluate the thermal conductivity,
          giving:
          \begin{equation}
            \Fb^\xv_{\mathrm{diffusive}}(\avg{\rho e}_{i+1/2,j}) = - \kth(\avg{\qb}_{i+1/2,j}) \left . \frac{\partial \avg{T}}{\partial x} \right |_{i+1/2,j}
          \end{equation}

      \end{enumerate}
      These derivatives are derived in the supplemental Jupyter notebook.


\end{itemize}

\subsubsection{Solving the reactive system}
\label{sec:fourth_burn}
To fourth-order, we need to concern ourselves with the difference
between cell centers and averages.  We want to solve
Eq.~\ref{eq:fourthorderupdate} for the updated cell-average state,
$\avg{\Uc}^{m+1,(k)}$, but we note that $\avg{\Rb(\Uc)} \ne
\Rb(\avg{\Uc})$ to fourth-order, so we can't solve this the same way
we do for the second-order method.  Our approach is to instead
solve a cell-center version first, and then use this to find 
the cell-average update.

To start, we compute  ${\bf C}$ at cell centers from the approximation
\begin{equation}
  {\bf C}_\ijk = \avg{\bf C}_\ijk - \frac{h^2}{24} \Delta^{(2)} \avg{\bf C}_\ijk.
\end{equation}
Then we solve 
\begin{equation}
  \Uc_\ijk^{m+1,(k+1)} - \delta t_m \Rbs{\Uc_\ijk^{m+1,(k+1)}} = \Uc_\ijk^{m,(k+1)} + \delta t_m {\bf C}_\ijk
\end{equation}
using the same techniques described above for second-order.  It is
tempting to then construct the final $\avg{\Uc}_\ijk$ by converting
$\Uc_\ijk$ to averages using the Laplacian, but this would break
conservation, since the advective flux difference is buried in
${\bf C}_\ijk$.  Instead, we use the $\Uc_\ijk$ to evaluate the instantaneous
reaction rates one more time, to construct $\Rb(\Uc_\ijk)$, and then
construct the average reactive source as:
\begin{equation}
  \avg{\Rbs{\Uc}}_\ijk = \Rbs{\Uc_\ijk} + \frac{h^2}{24} \Delta^{(2)} \Rbs{\Uc_\ijk}
\end{equation}
and finally, use this $\avg{\Rb(\Uc)}_\ijk$ in Eq.~\ref{eq:fourthorderupdate} to
get the final $\avg{\Uc}_\ijk$, as:
\begin{equation}
  \avg{\Uc}_\ijk^{{m+1},(k+1)} = \avg{\Uc}_\ijk^{m,(k+1)}
  + \delta t_m  \avg{\Rbs{\Uc}}_\ijk^{{m+1},(k+1)} + \delta t_m \avg{\bf C}_\ijk
\end{equation}

\subsubsection{Pure Hydrodynamics}

To test the integration scheme without reactions,
there is no nonlinear solve needed, and our system update
is purely explicit:
\begin{align}
      \avg{\Uc}_\ijk^{{m+1},(k+1)} = \avg{\Uc}_\ijk^{m+1,(k)}
            &+ \delta t_m \left [ \avg{\Advs{\Uc}}^{m,(k+1)}_\ijk - \avg{\Advs{\Uc}}^{m,(k)}_\ijk \right ] \nonumber \\
            &+ \Ic_m^{m+1}\left  ( \avg{\Advs{\Uc}}^{(k)}_\ijk \right )
\end{align}
This is straightforward to solve and is used to test our method.
By measuring the convergence of pure hydrodynamics problems, we can
assess whether our scheme gets fourth-order accuracy.

\section{Numerical Experiments}

\label{sec:tests}

We consider a number of different test problems here to assess the
behavior of the new SDC integration scheme.  We  use 3 different
solvers: the default corner transport upwind (CTU) piecewise parabolic method (PPM) solver \citep{ppmunsplit,
  millercolella:2002} in \castro\ with Strang split
reactions (we refer to this as Strang CTU), the second-order SDC
method with piecewise linear slope reconstruction in space (SDC-2), and the fourth-order SDC
method (SDC-4).  For the pure hydrodynamics problems, our focus is on
demonstrating that the overall fourth-order SDC algorithm converges
as expected, so we focus only on that solver.  We conclude by
demonstrating the SDC-4 method on two science problems: a burning
buoyant bubble and a flame.

Traditionally with Strang-splitting, we would instruct the ODE solver
that evolves the reactions to use a relatively tight tolerance,
resulting in many substeps for the integration of the reaction terms
over $\delta t$.  With the SDC implementation, we are using a fixed
number of temporal nodes, evaluating the reactions with our
hydrodynamics in a coupled integration.  So while the Strang-split
case uses a much tighter tolerance in integration the reactions, it is
solving the wrong equations very accurately (i.e., the uncoupled
system), while the SDC method solves the correct, coupled equations
with fixed integration points (and potentially less accurately).  We
explore the solution of several hydrodynamics and reactive flow
problems here to understand how the different approaches perform.

When setting the initial conditions for the fourth-order tests, we
first initialize the cell-centers, $\Uc_\ijk$, to the analytic initial
conditions and then convert from cell-centers to averages as
\begin{equation}
\avg{\Uc}_\ijk = \Uc_\ijk + \frac{h^2}{24} \Delta^{(2)} \Uc_\ijk
\end{equation}
We also note that all runs use slope limiters, which can impact the ability
to get ideal convergence behavior near discontinuities, but we choose this approach
since this is how the method would be run in scientific simulations.

Finally, for the SDC methods, the timestep is limited by
\begin{equation}
\delta t \le \mathcal{C} \min_\ijk \left \{ \left [ \sum_{d=1}^{D} \frac{|\Ub_\ijk \cdot {\bf e}_\mathrm{d}| + c_\ijk}{\Delta x_d} \right ]^{-1} \right\}
\end{equation}
where $D$ is the number of dimensions.  This is more restrictive in
multi-dimensions than the timestep constraint for
CTU~\citep{ppmunsplit}.  The dimensionless CFL number, $\mathcal{C}$
is kept less than 1 in our simulations, although we note that for
Runge-Kutta integration, \citet{mccorquodalecolella} suggest it can be
as high as 1.4.


\subsection{Gamma-Law Acoustic Pulse}

The gamma-law acoustic pulse problem is a pure hydrodynamics test.\deleted{We use
the initial conditions from \citet{mccorquodalecolella}}\footnote{This
  problem setup is available in \castro\ as {\tt
    Exec/hydro\_tests/acoustic\_pulse}.  The runs for the convergence test
  can be run using the {\tt convergence\_sdc4.sh} script there.}
This problem sets up a pressure perturbation in a 2D square domain in
a constant entropy background and watches the propagation of a sound
wave (as a ring) move outward from the perturbation.  The gamma-law
equation of state needs a composition to define the temperature (via
the ideal gas law), so we choose the composition to be pure
\isot{H}{1}.  
\added{The initial state is taken from \citet{mccorquodalecolella}:
\begin{equation}
\rho = \begin{cases}
 \rho_0 + f_\rho e^{-16r^2} \cos^6(\pi r) & r < 1/2 \\
 \rho_0 & r \ge 1/2
\end{cases} 
\end{equation}
with the pressure found assuming constant entropy:
\begin{equation}
p = \left ( \frac{\rho}{\rho_0} \right )^\gamma
\end{equation}
The parameters we use are given in Table~\ref{table:acoustic_pulse}.}
\begin{deluxetable}{lcc}
\tablecaption{\label{table:acoustic_pulse} Acoustic pulse parameters.}
\tablehead{\colhead{parameter} & \colhead{value}}
\startdata
$\rho_0$ & $1.4~\gcc$ \\
$f_\rho$    & $0.14~\gcc$ \\
$\gamma$ & $1.4$ \\
\enddata
\end{deluxetable}

\begin{figure}[t]
\centering
\plotone{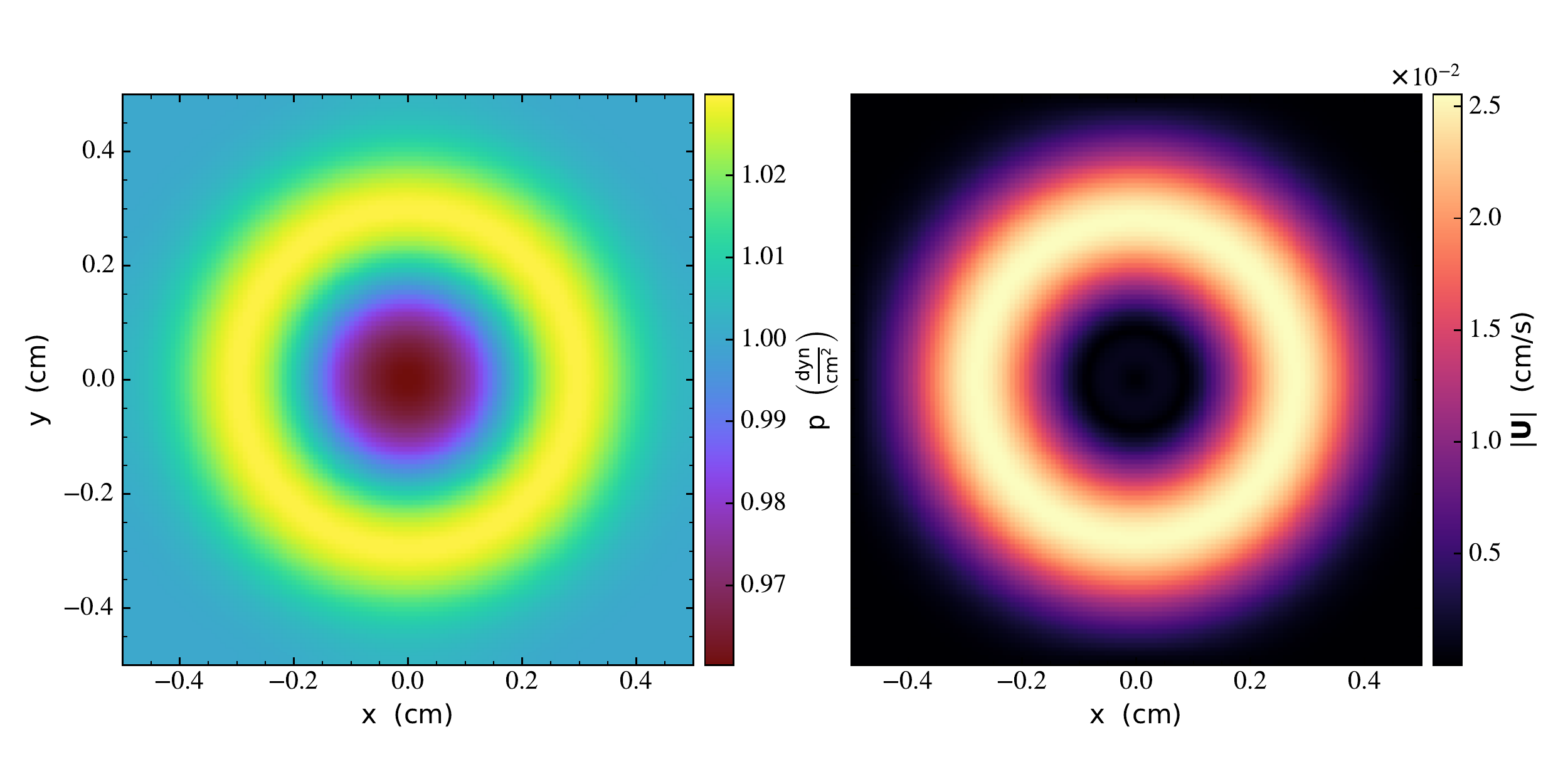}
\caption{\label{fig:acoustic_pulse} Pressure and velocity magnitude at
  $t = 0.24~\mathrm{s}$ for the acoustic pulse problem run with SDC-4
  using $128^2$ cells.}
\end{figure}

This test serves as a comparison to
the method in \citet{mccorquodalecolella}---we use the same fourth-order spatial
reconstruction, but use the SDC integration update instead of the
Runge-Kutta method used therein.  We run for 0.24~s using a
fixed timestep,
\begin{equation}
\delta t = 3\times 10^{-3} \left ( \frac{64}{n_\mathrm{zones}} \right )~\mathrm{s}
\end{equation}
on a domain $[0, 1]^2$ with periodic boundary conditions.
Figure~\ref{fig:acoustic_pulse} shows the state at the end of the
simulation.

We approximate the convergence rate by defining the error as the norm over
cells of the difference between a fine and coarse calculation,
differing by a factor of two\footnote{We use the \amrex\ {\tt
    RichardsonConvergenceTest} tool to compute the convergence rate
  (located in {\tt amrex/Tools/C\_util/Convergence}).}.  We run with
$64^2$, $128^2$, $256^2$ and $512^2$ cells, so $\epsilon_{64
  \rightarrow 128}$ is the error between the $64^2$ and $128^2$ cell
calculations.  We then estimate the convergence rate, $r$, from two
pairs of simulations, e.g., $r = \log_2 (\epsilon_{64 \rightarrow
  128}/\epsilon_{128 \rightarrow 256})$.
Table~\ref{table:acoustic_converge} shows the results in the $L_1$
norm, including the measured convergence rate.  We see fourth-order
convergence in all of the conserved variables and also in temperature.
This convergence agrees well with that presented in
\citet{mccorquodalecolella}.  We note this same test problem was also
used with SDC in \citet{Emmett:2018}.

\begin{deluxetable}{lllllll}
\tablecaption{\label{table:acoustic_converge} Convergence ($L_1$ norm) for the $\gamma$-law EOS
acoustic pulse problem using the SDC-4 solver.}
\tablehead{\colhead{field} & \colhead{$\epsilon_{64 \rightarrow 128}$} & 
           \colhead{rate} & \colhead{$\epsilon_{128\rightarrow 256}$} & 
           \colhead{rate} & \colhead{$\epsilon_{256\rightarrow 512}$}}
\startdata
 $\rho$                      & $3.625 \times 10^{-6}$  & 3.980  & $2.297 \times 10^{-7}$  & 3.995  & $1.441 \times 10^{-8}$  \\
 $\rho u$                    & $2.087 \times 10^{-6}$  & 3.969  & $1.332 \times 10^{-7}$  & 3.992  & $8.371 \times 10^{-9}$  \\
 $\rho v$                    & $2.087 \times 10^{-6}$  & 3.969  & $1.332 \times 10^{-7}$  & 3.992  & $8.371 \times 10^{-9}$  \\
 $\rho E$                    & $9.143 \times 10^{-6}$  & 3.980  & $5.794 \times 10^{-7}$  & 3.995  & $3.634 \times 10^{-8}$  \\
 $\rho e$                    & $9.093 \times 10^{-6}$  & 3.980  & $5.763 \times 10^{-7}$  & 3.995  & $3.614 \times 10^{-8}$  \\
 $T$                         & $8.855 \times 10^{-15}$ & 3.979  & $5.614 \times 10^{-16}$ & 3.995  & $3.521 \times 10^{-17}$ \\
\enddata
\end{deluxetable}

\subsection{Real Gas Acoustic Pulse}

\label{sec:real_gas_pulse}

\begin{figure}[t]
\centering
\plotone{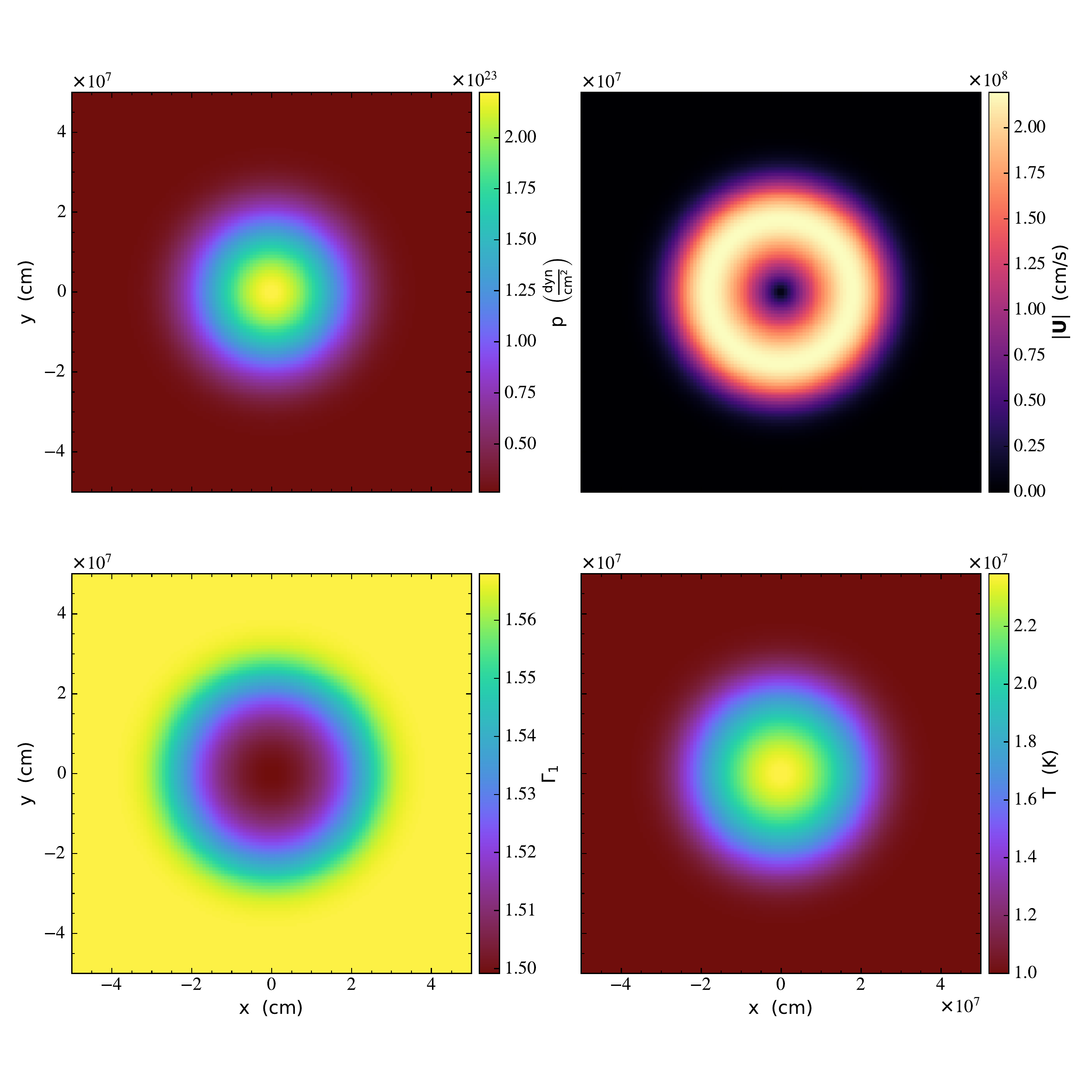}
\caption{\label{fig:acoustic_pulse_general} Pressure, velocity
  magnitude, $\Gamma_1$, and temperature at $t = 0.02~\mathrm{s}$ for
  the general EOS acoustic pulse problem run with SDC-4 using $128^2$
  cells.}
\end{figure}

To assess the performance with a real stellar EOS, we create a
generalized version of the acoustic pulse problem.  We use the
Helmholtz free energy based equation of state of
\cite{timmes_swesty:2000}, including degenerate/relativistic
electrons, ideal gas ions, and radiation.  Our initial conditions
are:
\begin{equation}
p = \begin{cases}
 p_0 \left [1 + f_p \, e^{-(r/\delta_r)^2} \cos^6(\pi r / L_x)\right ] & r < L_x/2 \\
 p_0 & r \ge L_x/2
\end{cases} 
\end{equation}
and 
\begin{equation}
s = s_0
\end{equation}
where $p_0$ and $s_0$ are the ambient pressure and specific entropy,
$\delta_r$ is the width of the perturbation, $f_p$ is the factor by
which pressure increases above ambient, $L_x$ is the physical width of
the domain in the $x$-direction, and $r$ is the distance from the
center of the domain.  We can then find the density and internal
energy from the equation of state\footnote{This problem setup is
  available in \castro\ as {\tt
    Exec/hydro\_tests/acoustic\_pulse\_general}.}.  Our equation of
state requires a composition---we make all of the material 
hydrogen ($A = Z = 1$).  We specify $p_0$ and $s_0$ in terms of
$\rho_0$ and $T_0$ using the equation of state, $p_0 = p(\rho_0, T_0)$
and $s_0 = s(\rho_0, T_0)$.  We run on a domain $[0, L_x]^2$, with
periodic boundaries, to a time of 0.02~s and use a fixed
timestep, scaled with resolution, $n_\mathrm{zones}$, as
\begin{equation}
\delta t = 2\times 10^{-4} \left (\frac{64}{n_\mathrm{zones}} \right )~\mathrm{s}
\end{equation}
Our choice of parameters is given in
Table~\ref{table:acoustic_general}.  These initial conditions were
picked to give a reasonable range of $\Gamma_1$ on the grid (it spans
$1.48$--$1.57$ initially).  We run this test for $64^2$, $128^2$, $256^2$,
and $512^2$ cells (in each direction).  We note that the amplitude of
our pressure perturbation is a bit large, and we have a Mach number of
$0.6$ at the end of the simulation---this suggests that the limiters
may have an effect here.  Figure~\ref{fig:acoustic_pulse_general} shows
the state after 0.02~s of evolution for the $128^2$ SDC-4 simulation.
Table~\ref{table:acoustic_general_converge} shows the convergence.  We
again see nearly fourth-order convergence for all flow variables.

\begin{deluxetable}{lcc}
\tablecaption{\label{table:acoustic_general} Stellar EOS acoustic pulse parameters.}
\tablehead{\colhead{parameter} & \colhead{value}}
\startdata
$\rho_0$ & $5\times 10^5~\gcc$ \\
$T_0$    & $10^7~\mathrm{K}$ \\
$f_p$    & $15$ \\
$\delta_r$ & $2\times 10^7~\mathrm{cm}$ \\
$L_x$  & $10^8~\mathrm{cm}$
\enddata
\end{deluxetable}

\begin{deluxetable}{lllllll}
\tablecaption{\label{table:acoustic_general_converge} Convergence ($L_1$ norm) for the real EOS
acoustic pulse problem using the SDC-4 solver.}
\tablehead{\colhead{field} & \colhead{$\epsilon_{64 \rightarrow 128}$} & 
           \colhead{rate} & \colhead{$\epsilon_{128\rightarrow 256}$} & 
           \colhead{rate} & \colhead{$\epsilon_{256\rightarrow 512}$}}
\startdata
 $\rho$                      & $1.935 \times 10^{17}$  & 3.939  & $1.262 \times 10^{16}$  & 3.981  & $7.988 \times 10^{14}$  \\
 $\rho u$                    & $3.842 \times 10^{25}$  & 3.907  & $2.562 \times 10^{24}$  & 3.972  & $1.633 \times 10^{23}$  \\
 $\rho v$                    & $3.842 \times 10^{25}$  & 3.907  & $2.562 \times 10^{24}$  & 3.972  & $1.633 \times 10^{23}$  \\
 $\rho E$                    & $4.079 \times 10^{34}$  & 3.939  & $2.659 \times 10^{33}$  & 3.981  & $1.684 \times 10^{32}$  \\
 $\rho e$                    & $3.526 \times 10^{34}$  & 3.949  & $2.283 \times 10^{33}$  & 3.982  & $1.444 \times 10^{32}$  \\
 $T$                         & $5.657 \times 10^{19}$  & 3.955  & $3.648 \times 10^{18}$  & 3.991  & $2.295 \times 10^{17}$  \\
\enddata
\end{deluxetable}

\subsection{Real Gas Shock Tubes}

In \cite{zingalekatz}, we examine exact solutions to shock tube
problems with the stellar equation of state to be used as test
problems for hydrodynamics schemes.  Here we run these same problems
with the Strang CTU and SDC-4 solvers. We
do not attempt to measure convergence here, since these problems
feature discontinuities, but instead run these to demonstrate that we
can recover the correct behavior for nonsmooth flows with a general equation of
state with the new fourth-order accurate solver.  

\begin{figure}[t]
\centering
\plotone{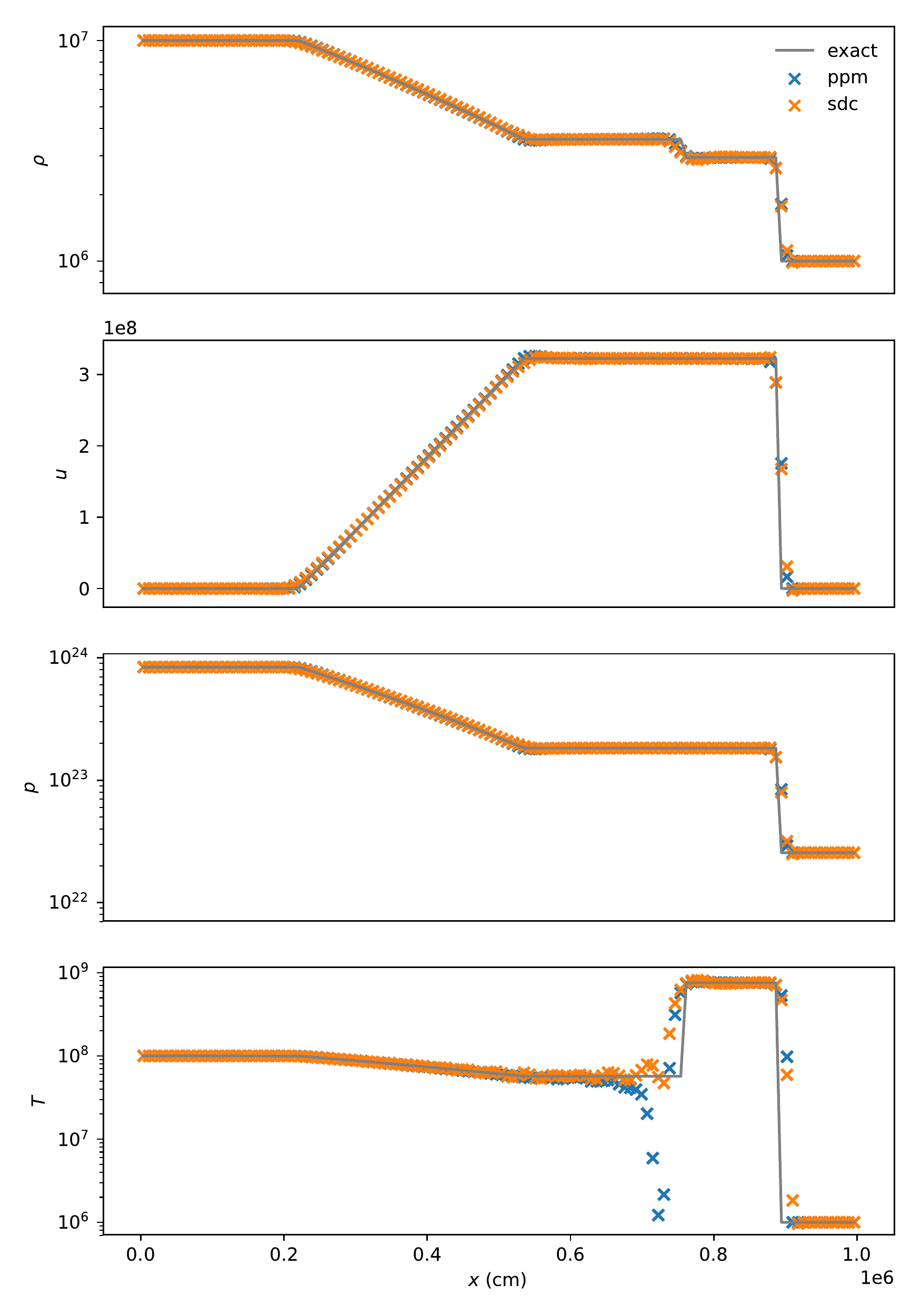}
\caption{\label{fig:test1} The stellar EOS Sod-like problem (test 1) from \cite{zingalekatz}.}
\end{figure}

\begin{figure}[t]
\centering
\plotone{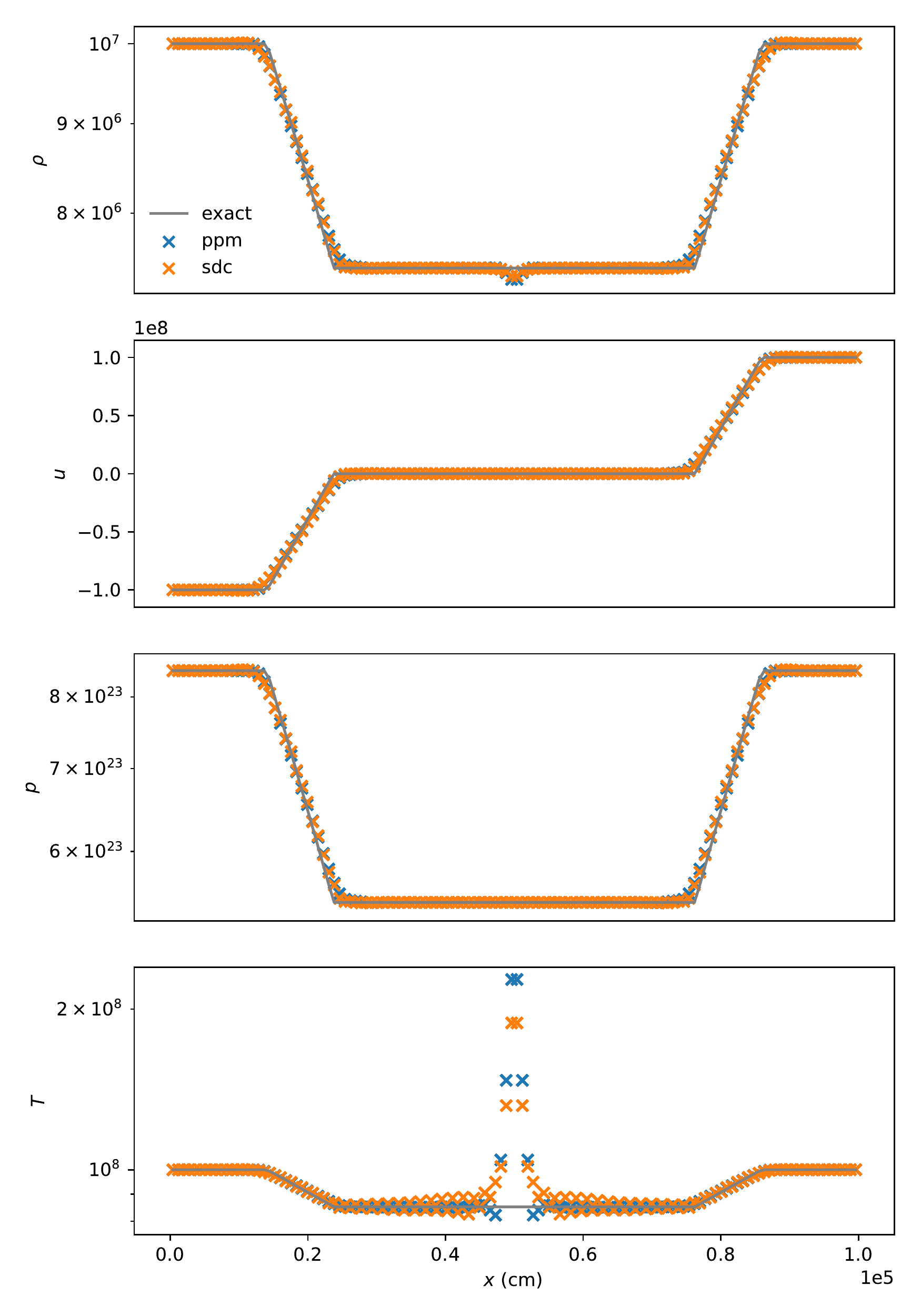}
\caption{\label{fig:test2} The stellar EOS double rarefaction problem (test 2) from \cite{zingalekatz}.}
\end{figure}

\begin{figure}[t]
\centering
\plotone{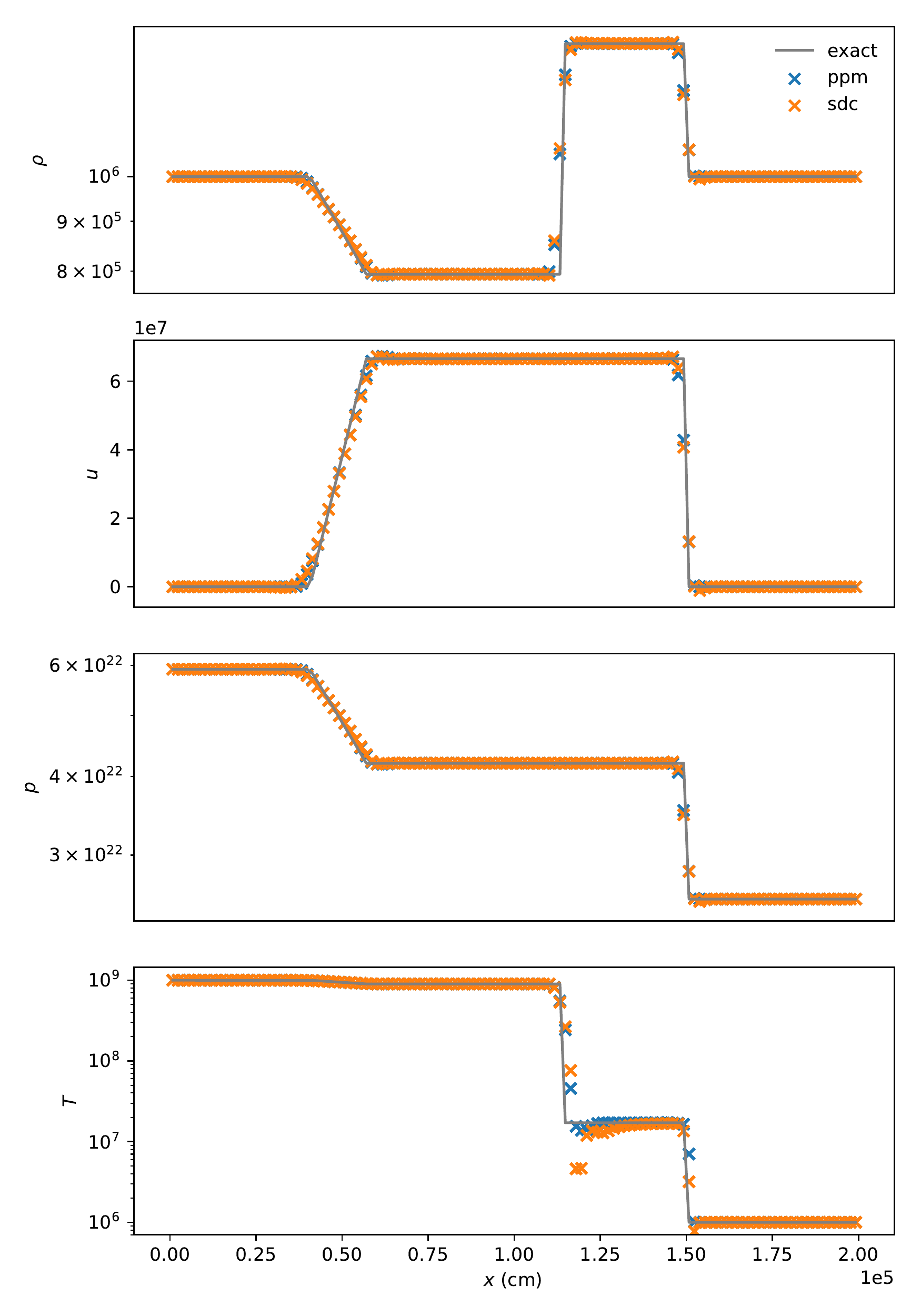}
\caption{\label{fig:test3} The stellar EOS strong shock problem (test 3) from \cite{zingalekatz}.}
\end{figure}

The first problem is a Sod-like problem (Figure~\ref{fig:test1}),
featuring a rightward moving shock and contact and a leftward moving
rarefaction.  The Strang CTU and SDC-4 solutions are shown together
with the exact solution.  We see that both solvers have trouble
with the temperature at the contact discontinuity (Strang CTU
undershoots while SDC-4 oscillates a bit), but otherwise the agreement
is quite good.  The second problem is a double rarefaction
(Figure~\ref{fig:test2}). The initial thermodynamic state is constant
but with outward directed velocities at the interface.  A vacuum
region forms in-between two rarefactions.  Both the Strang CTU and
SDC-4 method have difficultly with the temperature at the very center
(where both $p$ and $\rho$ are going to zero), but otherwise agree
nicely with the analytic solution.  The final problem is a strong
shock (Figure~\ref{fig:test3}).  Again both methods have difficulty
with the temperature at the contact discontinuity with the SDC-4
solution undershooting a bit more than the Strang CTU solution.
Overall, these tests show that for problems involving shocks, our
fourth-order scheme is working as expected.

\subsection{Thermal diffusion test}

The standard test problem for thermal diffusion is to diffuse a
Gaussian temperature profile with a constant diffusion coefficient,
which remains Gaussian but with a lower amplitude and greater width as
time evolves.  However, we want to ensure we converge properly for a
state-dependent conductivity.  To test this, we use a simple powerlaw
thermal conductivity:
\begin{equation}
  \kth = {\kth}_0 T^\nu
\end{equation}
We adopt ${\kth}_0 = 1$ and $\nu = 2$.  We still begin with a Gaussian
profile of the form:
\begin{equation}
  T(r) = T_1 + (T_2 - T_1) e^{-r^2/(4\mathcal{D}t_0)}
\end{equation}
where $r$ is the distance from the center of the domain, $\mathcal{D}$ is the thermal diffusivity,
\begin{equation}
  \mathcal{D} = \frac{\kth}{\rho c_v}
\end{equation}
and $t_0$ has units of time and serves to control the initial width of
the Gaussian.  We take $t_0 = 10^{-3}~\mathrm{s}$ here, and turn
off hydrodynamics, so only the temperature and internal energy 
evolve in this test.  We use a gamma-law equation of state and a pure
hydrogen composition (with $\gamma = 5/3$), so the specific heat is
just
\begin{equation}
  c_v = \frac{3}{2} \frac{k_B}{m_u}
\end{equation}
where $k_B$ is Boltzmann's constant and $m_u$ is the atomic mass unit.
We choose the constant density in the domain, $\rho_0$, so that the
thermal diffusivity in the center is $\mathcal{D}(r=0) = 1$.
This gives:
\begin{equation}
  \rho_0 = \frac{\kth(T_2)}{c_v(T_2)}
\end{equation}

Finally, we use the standard explicit diffusion timestep limiter, of the form:
\begin{equation}
\label{eq:difflimit}
\delta t_\mathrm{diff} = \frac{\mathcal{C}}{2}\min \left \{ \frac{\Delta x^2}{\mathcal{D}} \right  \}
\end{equation}
where we use the same CFL factor as with hydrodynamics to reduce the timestep.

\begin{figure}[t]
\centering
\plotone{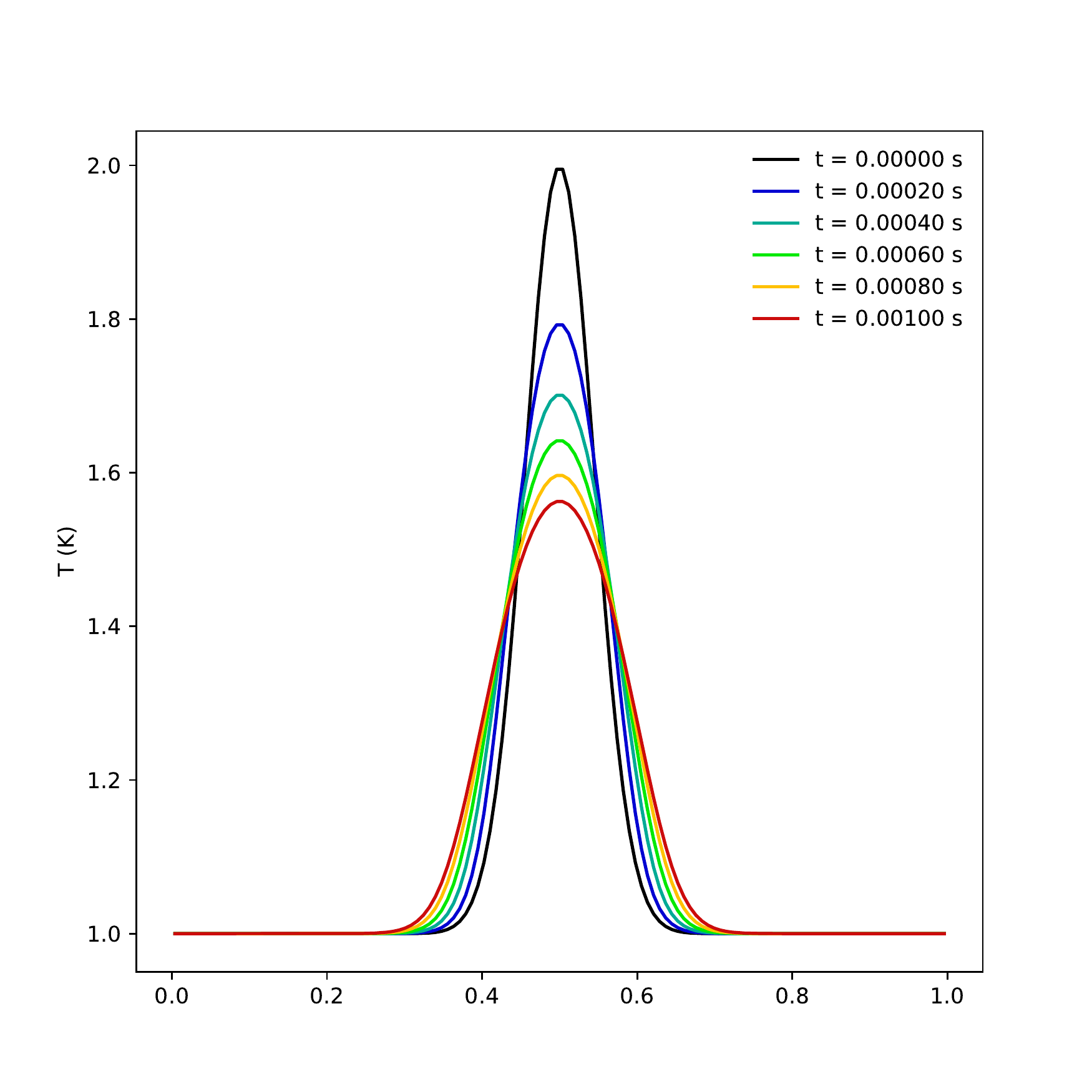}
\caption{\label{fig:diffusion_1d} Temperature profiles for the 1-d thermal diffusion
test run with 128 zones.}
\end{figure}

\begin{deluxetable}{lllllll}
\tablecaption{\label{table:diffusion_test_1d} Convergence ($L_1$ norm)
  for the 1-d and 2-d thermal diffusion test with fourth-order SDC.}
\tablehead{\colhead{field} & \colhead{$\epsilon_{64 \rightarrow 128}$} & 
           \colhead{rate} & \colhead{$\epsilon_{128\rightarrow 256}$} & 
           \colhead{rate} & \colhead{$\epsilon_{256\rightarrow 512}$}}
\startdata
\cutinhead{1-d test}
 $\rho e$                    & $1.112 \times 10^{-5}$  & 3.949  & $7.198 \times 10^{-7}$  & 3.987  & $4.539 \times 10^{-8}$  \\
 $T$                         & $1.063 \times 10^{-5}$  & 3.953  & $6.867 \times 10^{-7}$  & 3.975  & $4.368 \times 10^{-8}$  \\
\cutinhead{2-d test}
 $\rho e$                    & $1.902 \times 10^{-6}$  & 3.958  & $1.224 \times 10^{-7}$  & 3.987  & $7.719 \times 10^{-9}$  \\
 $T$                         & $1.770 \times 10^{-6}$  & 3.966  & $1.133 \times 10^{-7}$  & 3.991  & $7.127 \times 10^{-9}$  \\
\enddata
\end{deluxetable}

We run in 1-d on a domain $[0, 1]$, with 64, 128, 256, and 512 cells
for $10^{-3}~\mathrm{s}$, with $\mathcal{C} = 0.5$.
Figure~\ref{fig:diffusion_1d} shows the temperature profile at various
times.  Table~\ref{table:diffusion_test_1d} shows the convergence for
the test in 1-d---we see nearly perfect fourth-order convergence.  We
also run in 2-d on $[0, 1]^2$ with $64^2$, $128^2$, $256^2$, and
$512^2$ cells for the same time.  In 2-d, we exercise the
face-averaging of the diffusive fluxes.  The same table shows the
convergence for 2-d, and again we see nearly perfect fourth-order
convergence.

\subsection{Reacting Hydrodynamics Test}

Next we adapt the general EOS acoustic pulse problem from section \ref{sec:real_gas_pulse}
to include reactions, which enables us to test the convergence
rate of the coupled hydrodynamics and reactions update.  The problem setup is
the same, but we now initialize the material to be completely
\isot{He}{4} and we use a simple reaction network with the
triple-alpha and $\isotm{C}{12}(\alpha,\gamma)\isotm{O}{16}$
reactions\footnote{This problem setup is available in \castro\ as {\tt
    Exec/reacting\_tests/reacting\_convergence}}, using rates from
\citet{caughlan-fowler:1988} along with screening from
\citet{graboske:1973,alastuey:1978,itoh:1979}.  The network also contains \isot{Fe}{56}, which is not linked to any other nuclei via reactions (it is used as an inert marker).  This network is available as part of the 
StarKiller microphysics project~\citep{starkiller_19_08}.  Since we start out as
\isot{He}{4}, any \isot{C}{12} or \isot{O}{16} in the final output is
created via the nucleosynthesis, so these species can help understand
the convergence of the reactions.  

For all the SDC runs, we use the simple Newton solve, the
analytic estimate of the Jacobian, solve for $(\rho e)$ in the update,
and set the tolerances as $\epsrho = 10^{-10}$, $\epsspec = 10^{-10}$,
$\epsener = 10^{-5}$, and $\epsabs = 10^{-10}$. 

We run with the same timestep as the non-reacting version to a stop
time of 0.06~s. Here we compute the convergence rate for the
Strang CTU, SDC-2, and SDC-4 solvers.   All simulations are
run in 2-d.

\begin{figure}[t]
\centering
\plotone{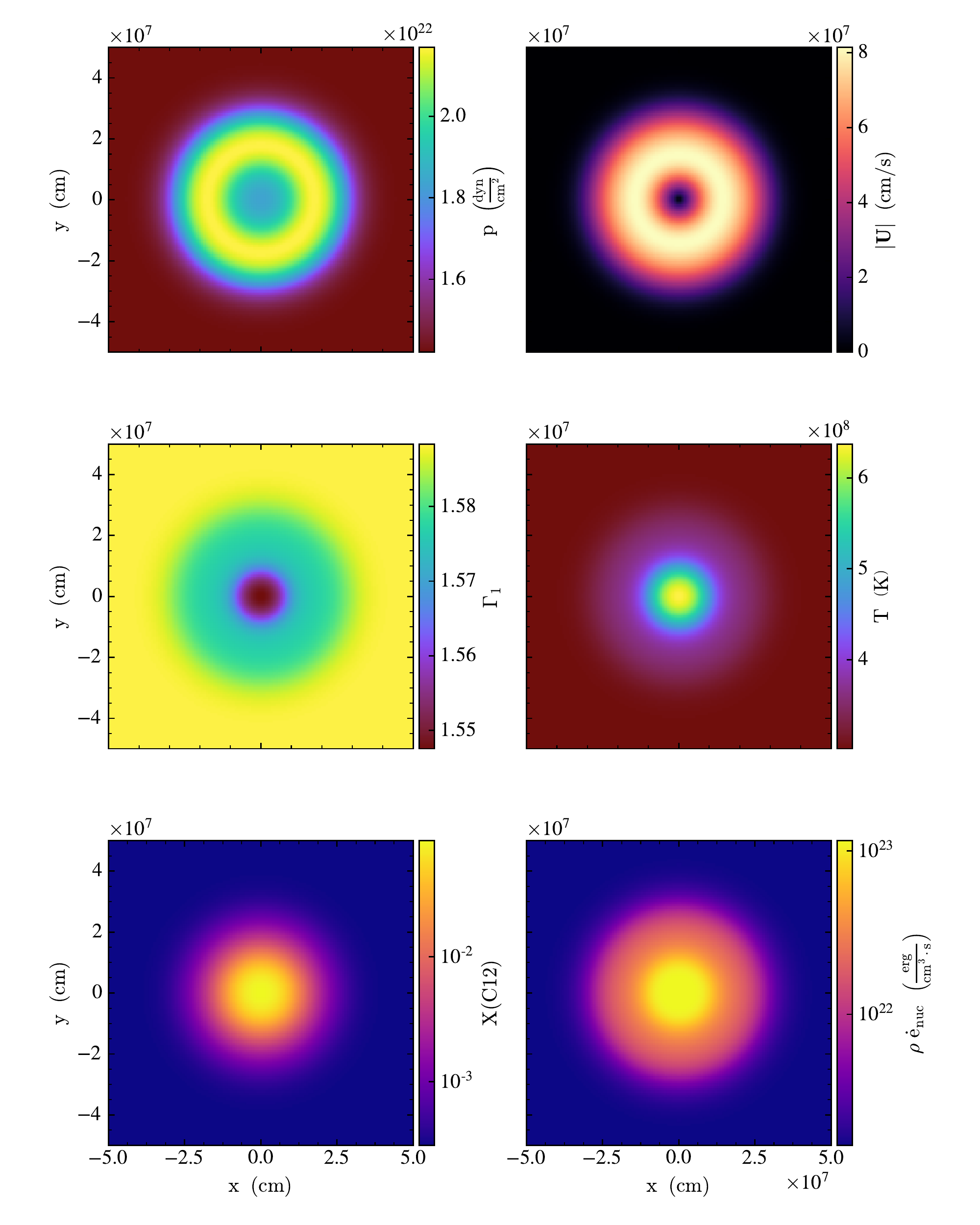}
\caption{\label{fig:reacting_convergence} Thermodynamic, dynamic,
  and nuclear state at 
  $t = 0.06~\mathrm{s}$ for the reacting hydrodynamics test run with SDC-4
  using $128^2$ cells.}
\end{figure}

\begin{deluxetable}{lllllll}
\tablecaption{\label{table:reacting_convergence_strang} Convergence
  ($L_1$ norm) for the reacting convergence problem with the  Strang
  CTU solver.}
\tablehead{\colhead{field} & \colhead{$\epsilon_{64 \rightarrow 128}$} & 
           \colhead{rate} & \colhead{$\epsilon_{128\rightarrow 256}$} & 
           \colhead{rate} & \colhead{$\epsilon_{256\rightarrow 512}$}}
\startdata
 $\rho$                      & $2.780 \times 10^{18}$  & 2.051  & $6.706 \times 10^{17}$  & 2.580  & $1.121 \times 10^{17}$  \\
 $\rho u$                    & $6.780 \times 10^{26}$  & 2.446  & $1.245 \times 10^{26}$  & 2.907  & $1.659 \times 10^{25}$  \\
 $\rho v$                    & $6.780 \times 10^{26}$  & 2.446  & $1.245 \times 10^{26}$  & 2.907  & $1.659 \times 10^{25}$  \\
 $\rho E$                    & $2.465 \times 10^{35}$  & 2.333  & $4.893 \times 10^{34}$  & 2.650  & $7.797 \times 10^{33}$  \\
 $\rho e$                    & $2.268 \times 10^{35}$  & 2.298  & $4.611 \times 10^{34}$  & 2.721  & $6.991 \times 10^{33}$  \\
 $T$                         & $2.245 \times 10^{21}$  & 1.682  & $6.995 \times 10^{20}$  & 2.439  & $1.290 \times 10^{20}$  \\
 $\rho X(\isotm{He}{4})$     & $2.861 \times 10^{18}$  & 2.027  & $7.018 \times 10^{17}$  & 2.553  & $1.195 \times 10^{17}$  \\
 $\rho X(\isotm{C}{12})$     & $1.717 \times 10^{17}$  & 1.945  & $4.458 \times 10^{16}$  & 2.194  & $9.745 \times 10^{15}$  \\
 $\rho X(\isotm{O}{16})$     & $1.717 \times 10^{14}$  & 1.648  & $5.479 \times 10^{13}$  & 1.898  & $1.471 \times 10^{13}$  \\
 $\rho X(\isotm{Fe}{56})$    & $2.780 \times 10^{-12}$ & 2.051  & $6.706 \times 10^{-13}$ & 2.580  & $1.121 \times 10^{-13}$ \\
\enddata
\end{deluxetable}

\begin{deluxetable}{lllllll}
\tablecaption{\label{table:reacting_convergence_sdc} Convergence ($L_1$ norm) for the reacting convergence problem with the SDC-2 solver.}
\tablehead{\colhead{field} & \colhead{$\epsilon_{64 \rightarrow 128}$} & 
           \colhead{rate} & \colhead{$\epsilon_{128\rightarrow 256}$} & 
           \colhead{rate} & \colhead{$\epsilon_{256\rightarrow 512}$}}
\startdata
 $\rho$                      & $2.024 \times 10^{18}$  & 2.011  & $5.022 \times 10^{17}$  & 2.021  & $1.238 \times 10^{17}$  \\
 $\rho u$                    & $3.720 \times 10^{26}$  & 2.063  & $8.901 \times 10^{25}$  & 2.030  & $2.180 \times 10^{25}$  \\
 $\rho v$                    & $3.720 \times 10^{26}$  & 2.063  & $8.901 \times 10^{25}$  & 2.030  & $2.180 \times 10^{25}$  \\
 $\rho E$                    & $2.302 \times 10^{35}$  & 2.030  & $5.635 \times 10^{34}$  & 2.014  & $1.395 \times 10^{34}$  \\
 $\rho e$                    & $2.053 \times 10^{35}$  & 2.025  & $5.043 \times 10^{34}$  & 2.013  & $1.249 \times 10^{34}$  \\
 $T$                         & $1.643 \times 10^{21}$  & 2.060  & $3.939 \times 10^{20}$  & 2.026  & $9.676 \times 10^{19}$  \\
 $\rho X(\isotm{He}{4})$     & $2.002 \times 10^{18}$  & 2.015  & $4.951 \times 10^{17}$  & 2.027  & $1.215 \times 10^{17}$  \\
 $\rho X(\isotm{C}{12})$     & $1.042 \times 10^{17}$  & 2.032  & $2.546 \times 10^{16}$  & 2.019  & $6.281 \times 10^{15}$  \\
 $\rho X(\isotm{O}{16})$     & $1.564 \times 10^{14}$  & 1.935  & $4.090 \times 10^{13}$  & 2.003  & $1.020 \times 10^{13}$  \\
 $\rho X(\isotm{Fe}{56})$    & $2.024 \times 10^{-12}$ & 2.011  & $5.022 \times 10^{-13}$ & 2.021  & $1.238 \times 10^{-13}$ \\
\enddata
\end{deluxetable}

\begin{deluxetable}{lllllll}
\tablecaption{\label{table:reacting_convergence_sdc4} Convergence ($L_1$ norm) for the reacting convergence problem with the SDC-4 solver.}
\tablehead{\colhead{field} & \colhead{$\epsilon_{64 \rightarrow 128}$} & 
           \colhead{rate} & \colhead{$\epsilon_{128\rightarrow 256}$} & 
           \colhead{rate} & \colhead{$\epsilon_{256\rightarrow 512}$}}
\startdata
 $\rho$                      & $2.127 \times 10^{17}$  & 3.855  & $1.470 \times 10^{16}$  & 3.972  & $9.369 \times 10^{14}$  \\
 $\rho u$                    & $3.401 \times 10^{25}$  & 3.856  & $2.349 \times 10^{24}$  & 3.958  & $1.511 \times 10^{23}$  \\
 $\rho v$                    & $3.401 \times 10^{25}$  & 3.856  & $2.349 \times 10^{24}$  & 3.958  & $1.511 \times 10^{23}$  \\
 $\rho E$                    & $1.945 \times 10^{34}$  & 3.891  & $1.311 \times 10^{33}$  & 3.953  & $8.463 \times 10^{31}$  \\
 $\rho e$                    & $1.672 \times 10^{34}$  & 3.899  & $1.120 \times 10^{33}$  & 3.955  & $7.223 \times 10^{31}$  \\
 $T$                         & $1.236 \times 10^{20}$  & 3.708  & $9.463 \times 10^{18}$  & 3.949  & $6.125 \times 10^{17}$  \\
 $\rho X(\isotm{He}{4})$     & $2.147 \times 10^{17}$  & 3.858  & $1.481 \times 10^{16}$  & 3.969  & $9.458 \times 10^{14}$  \\
 $\rho X(\isotm{C}{12})$     & $8.789 \times 10^{15}$  & 3.798  & $6.319 \times 10^{14}$  & 3.911  & $4.201 \times 10^{13}$  \\
 $\rho X(\isotm{O}{16})$     & $1.294 \times 10^{13}$  & 3.765  & $9.518 \times 10^{11}$  & 3.872  & $6.501 \times 10^{10}$  \\
 $\rho X(\isotm{Fe}{56})$    & $2.127 \times 10^{-13}$ & 3.855  & $1.470 \times 10^{-14}$ & 3.972  & $9.369 \times 10^{-16}$ \\
\enddata
\end{deluxetable}

Figure~\ref{fig:reacting_convergence} shows the thermodynamic, dynamic, and nuclear state
for the $128^2$ SDC-4 simulation at 0.06~s. 
 We run with $64^2$, $128^2$, $256^2$, and $512^2$ cells
and compute the error between successive resolutions and measure the
convergence rate.  Tables~\ref{table:reacting_convergence_strang},
\ref{table:reacting_convergence_sdc}, and
\ref{table:reacting_convergence_sdc4} show the convergence.  We see
that the Strang CTU algorithm achieves second order for most variables
(as expected), with some quantities converging almost third order (for
smooth flows, the PPM algorithm approaches third order accuracy in
space), while having difficulty with \isot{O}{16}.  For SDC-2, we see
second-order convergence in all the variables, including \isot{O}{16}.
Finally, for SDC-4, all of the variables converge at rates of
$\sim3.8$--$3.9$, demonstrating the fourth-order accuracy expected for the
method.  This test shows that the SDC algorithm can achieve
fourth-order convergence for reactive hydrodynamics problems with
astrophysical networks.

\subsection{Burning Buoyant Bubble}

Our final convergence test problem considers a hydrostatic atmosphere with a
temperature perturbuation\footnote{This problem setup is available in \castro\ as {\tt Exec/reacting\_tests/bubble\_convergence}.}.  Buoyancy causes the perturbation to rise (and
eventually roll up in the nonlinear phase).  The presence of reactions
prevents the bubble from fizzling out, keeping it buoyant via the
heat deposition.  In addition to looking at the convergence rate  of the
numerical solutions, we also consider how well we maintain hydrostatic
equilibrium in an undisturbed hydrostatic atmosphere.

We create an initial atmospheric model that is isentropic and in hydrostatic equilibrium by integrating
the system:
\begin{align}
  \frac{dp}{dy} &= -\rho(p, s) g \\
  \frac{ds}{dy} &= 0 
\end{align}
where the relation $\rho(p, s)$ is provided by our equation of state.
We take gravity, $g$, to be constant and the composition to be uniform
throughout the atmosphere (pure helium, with the other nuclei mass
fractions set to the small value $10^{-8}$).  To integrate this
system, we specify the conditions at the base of the atmosphere, which
we take to be the lower domain boundary (not the center of the
bottom-most cell).  We specify $\rho_\base$, and $T_\base$ and get
$p_\base$ and $s_\base$ through the general stellar equation of state.
We integrate this system using fourth-order Runge-Kutta, using a step
size of $\Delta x/2$ to get from the bottom of the domain to the first
cell-center, and then a step size of $\Delta x$ to integrate to each
of the remaining cell-centers vertically in the domain.  The initial
conditions are then converted to cell-averages using the same
transformation discussed earlier in the paper.  Note: the hydrostatic
model is generated specifically for the resolution of the problem, and
as such, the initial atmosphere converges with fourth-order accuracy.
For the boundary conditions, we use periodic conditions on the sides and
reflecting boundary conditions at the top and bottom.
Table~\ref{table:bbb_params} lists the problem setup parameters.

\begin{deluxetable}{lcc}
\tablecaption{\label{table:bbb_params} Hydrostatic atmosphere initial condition parameters.}
\tablehead{\colhead{parameter} & \colhead{value}}
\startdata
$\rho_\base$                & $10^7~\gcc$ \\
$T_\base$                   & $10^8~\mathrm{K}$ \\
$g$                         & $10^{10}~\mathrm{cm~s^{-2}}$ \\
$L_x = L_y$ (domain size)   & $7.68\times 10^6~\mathrm{cm}$ \\
$\sigma$                    & $2.56\times 10^5~\mathrm{cm}$ \\
\enddata
\end{deluxetable}

To test this initial setup, we evolve just the hydrostatic atmosphere
on our 2-d grid.  Analytically, the velocity should remain zero, if
hydrostatic equilibrium cancellation were perfect.  Due to truncation
error, a velocity does  build up over time, so we use the maximum of
the velocity magnitude, $|\Ub|$, as the measure of the error.  
Table~\ref{table:hse_error} lists this error for several resolutions.
We note that the velocity magnitudes are quite small, and we also see
fourth-order convergence as we increase the resolution.  This suggests
that with the fourth-order method, we can accurately maintain an atmosphere
in HSE without the need for well-balanced schemes \citep{ppm-hse,kappeli:2016}.

\begin{deluxetable}{lllll}
\tablecaption{\label{table:hse_error} Convergence of $\max\{|\Ub|\}$ for 
  an unperturbed hydrostatic atmosphere with fourth-order SDC.}
\tablehead{\colhead{$64^2$} & 
           \colhead{rate} & \colhead{$128^2$} & 
           \colhead{rate} & \colhead{$256^2$}}
\startdata
$1.884\times 10^{-2}$  & 3.987  & $1.188 \times 10^{-3}$ & 3.825 & $8.383\times 10^{-5}$ \\
\enddata
\end{deluxetable}

Next we add a perturbation and enable reactions, using the same
$3$-$\alpha$ + $\isotm{C}{12}(\alpha,\gamma)\isotm{O}{16}$ network
described above.  To perturb the atmosphere, we modify the temperature as:
\begin{equation}
T(x, y) = T_0(y) \left \{ 1 + \frac{3}{5} \left [ 1 + \tanh(4 - r) \right ] \right \}
\end{equation}
where $T_0$ is the temperature of the initial hydrostatic atmosphere
at the height $y$, and $r$ is the distance from the center of the domain.  The amplitude of the
perturbation was chosen to give a reasonable amount of burning to
\isot{C}{12} while keeping the Mach number below 0.1, while the shape was chosen to
give a flat central region.  We then recompute
the pressure at each point in the atmosphere through the equation of state,
constraining it to the hydrostatic pressure at the altitude, $p(y)$:
\begin{equation}
\rho(x, y) = \rho(T(x, y), p(y))
\end{equation}
This reduces the density, creating the initial buoyancy.  We run on
domains $64^2$, $128^2$, $256^2$, and $512^2$, to
$0.1~\mathrm{s}$ using a fixed timestep:
\begin{equation}
\delta t = 1.5\times 10^{-4} \left ( \frac{64}{n_\mathrm{zones}} \right )~\mathrm{s}.
\end{equation}
This is a difficult test problem because of the extreme nonlinearily of 
the dynamics.  The end time is picked so we measure convergence before the
bubble begins to roll-up in a strongly nonlinear fashion.  If we ran
longer, the strong temperature dependence of the $3$-$\alpha$ burning
would give strong nonlinear energy generation from local hot spots,
making a convergence test difficult for the lowest resolution
simulations we consider here.

\begin{figure}
\centering
\plotone{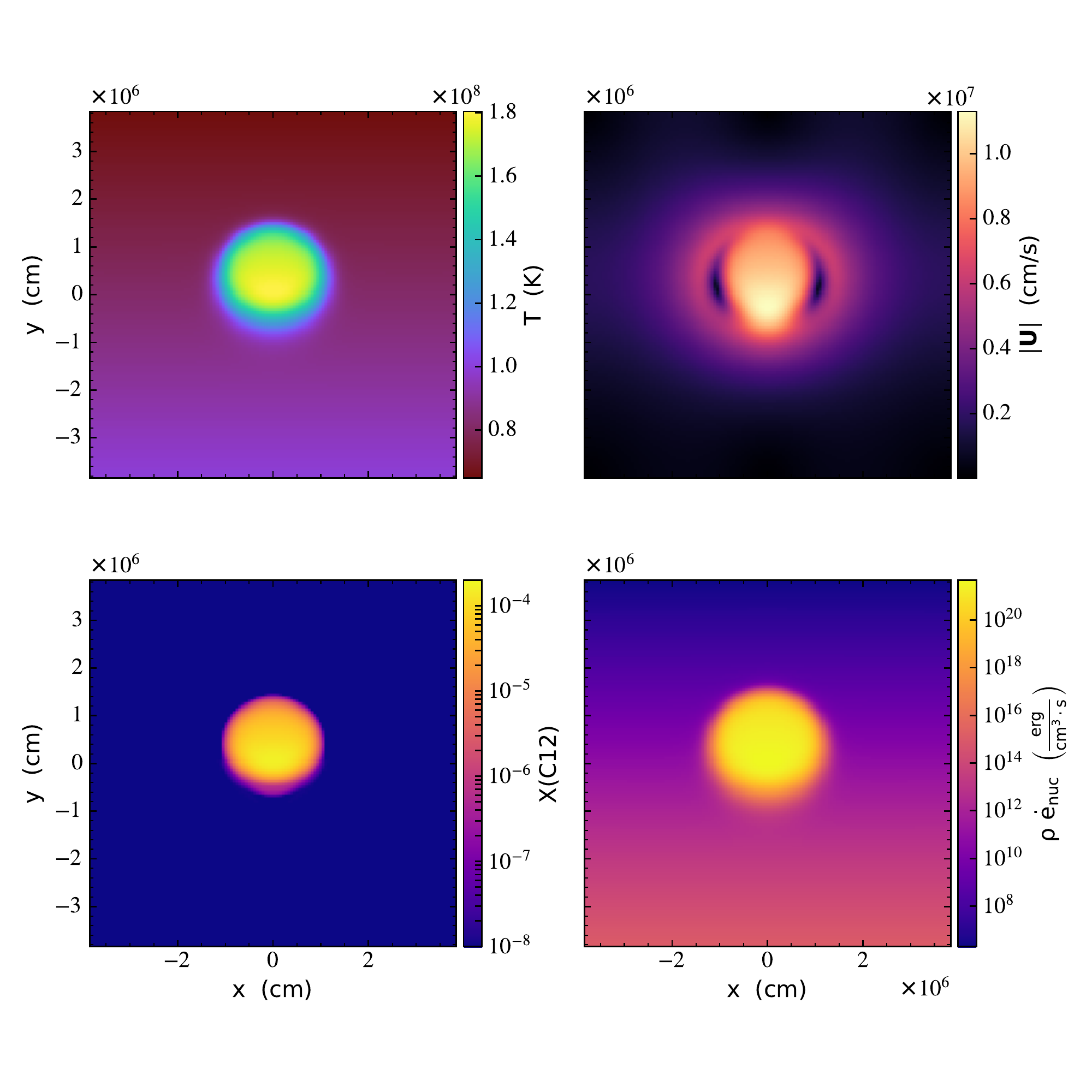}
\caption{\label{fig:bbb} Final state of the burning buoyant bubble problem for the $128^2$ simulation.}
\end{figure}

Figure~\ref{fig:bbb} shows the state of the bubble at the end point.
Table~\ref{table:bbb} shows the convergence across these problem
sizes.  At the lowest resolution, we barely resolve the burning
region, which affects the convergence, but we see nearly fourth-order
convergence for the higher resolution simulations.  Again, this test demonstrates
our SDC-4 method works as expected.

\begin{deluxetable}{lllllll}
\tablecaption{\label{table:bbb} Convergence ($L_1$ norm) for the burning buoyant bubble problem
using the SDC-4 solver.}
\tablehead{\colhead{field} & \colhead{$\epsilon_{64 \rightarrow 128}$} & 
           \colhead{rate} & \colhead{$\epsilon_{128\rightarrow 256}$} & 
           \colhead{rate} & \colhead{$\epsilon_{256\rightarrow 512}$}}
\startdata
 $\rho$                      & $3.591 \times 10^{15}$  & 3.263  & $3.739 \times 10^{14}$  & 3.713  & $2.852 \times 10^{13}$  \\
 $\rho u$                    & $1.120 \times 10^{24}$  & 3.794  & $8.072 \times 10^{22}$  & 3.930  & $5.296 \times 10^{21}$  \\
 $\rho v$                    & $1.314 \times 10^{24}$  & 3.544  & $1.127 \times 10^{23}$  & 3.838  & $7.879 \times 10^{21}$  \\
 $\rho E$                    & $3.701 \times 10^{32}$  & 2.946  & $4.801 \times 10^{31}$  & 3.647  & $3.834 \times 10^{30}$  \\
 $\rho e$                    & $3.701 \times 10^{32}$  & 2.946  & $4.801 \times 10^{31}$  & 3.646  & $3.834 \times 10^{30}$  \\
 $T$                         & $1.438 \times 10^{18}$  & 3.508  & $1.264 \times 10^{17}$  & 3.829  & $8.899 \times 10^{15}$  \\
 $\rho X(\isotm{He}{4})$     & $3.589 \times 10^{15}$  & 3.266  & $3.732 \times 10^{14}$  & 3.711  & $2.850 \times 10^{13}$  \\
 $\rho X(\isotm{C}{12})$     & $1.520 \times 10^{13}$  & 2.544  & $2.606 \times 10^{12}$  & 3.797  & $1.874 \times 10^{11}$  \\
 $\rho X(\isotm{O}{16})$     & $3.589 \times 10^{7}$   & 3.262  & $3.742 \times 10^{6}$   & 3.714  & $2.851 \times 10^{5}$   \\
 $\rho X(\isotm{Fe}{56})$    & $3.590 \times 10^{7}$   & 3.263  & $3.739 \times 10^{6}$   & 3.713  & $2.852 \times 10^{5}$   \\
\enddata
\end{deluxetable}

\subsection{Proof-of-concept: Helium deflagration}

To demonstrate that the SDC methods work with more extensive networks, we
run a 1-d helium deflagration with a 13 isotope alpha network, using
conditions that are appropriate to an sub-Chandra model of Type Ia
supernovae\footnote{This problem setup is available in \castro\ as
  {\tt Exec/science/flame}.}.  We do not try to assess convergence of
the flame here, because of the large number of timesteps ($\sim 10^6$)
needed to get ignition and the extreme nonlinearity of the burning.
Instead, we wish to demonstrate that the SDC-4 method can evolve a
flame using only the simple Newton iterations for the solution instead
of needing to solve an ODE system as we do with Strang-split methods.

We start by defining a fuel state in terms of density, temperature,
and composition: $\rho_\mathrm{fuel}$, $T_\mathrm{fuel}$,
$X_\mathrm{fuel}$.  From these conditions, we define an ambient
pressure through the equation of state:
\begin{equation}
  p_\mathrm{ambient} = p(\rho_\mathrm{fuel}, T_\mathrm{fuel}, X_\mathrm{fuel})
\end{equation}
We keep the pressure constant throughout the domain.  We then
define an ash temperature and composition, $T_\mathrm{ash}$ and
$X_\mathrm{ash}$, and we smoothly transition from the fuel to ash state
as:
\begin{align}
  T &= T_\mathrm{fuel} + \frac{1}{2} (T_\mathrm{ash} - T_\mathrm{fuel}) \left [ 1 - \tanh \left ( \frac{x - x_\mathrm{int}}{\delta_\mathrm{blend}} \right ) \right ] \\
  X_k &= X_{\mathrm{fuel},k} + \frac{1}{2} (X_{\mathrm{ash},k} - X_{\mathrm{fuel},k}) \left [ 1 - \tanh \left ( \frac{x - x_\mathrm{int}}{\delta_\mathrm{blend}} \right ) \right ]
\end{align}
and find the ash density by constraining the
conditions to be isobaric with the fuel through the equation of state:
\begin{equation}
  \rho_\mathrm{ash} = \rho(p_\mathrm{ambient}, T_\mathrm{ash}, X_\mathrm{ash})
\end{equation}
Here, $x_\mathrm{int}$ is the location of the initial transition
between ash (on the left) and fuel (on the right), and
$\delta_\mathrm{blend}$ is the width of the transition.  The remaining
thermodynamic quantities are found via the equation of state.  We use
a domain $[0, L_x]$ with simple zero-gradient boundary conditions.  The parameters used
for our simulation are shown in Table~\ref{table:he_flame}.

\begin{deluxetable}{lcc}
\tablecaption{\label{table:he_flame} Helium flame initial condition parameters.}
\tablehead{\colhead{parameter} & \colhead{value}}
\startdata
$\rho_\mathrm{fuel}$ & $2\times 10^7~\gcc$ \\
$T_\mathrm{fuel}$    & $5\times 10^7~\mathrm{K}$ \\
$X_\mathrm{fuel}(\isotm{He}{4})$ & 1.0 \\
$T_\mathrm{ash}$     & $3.6\times 10^9~\mathrm{K}$ \\
$X_\mathrm{ash}(\isotm{Ni}{56})$ & 1.0 \\
$L_x$                & $256~\mathrm{cm}$ \\
$x_\mathrm{int}$     & $0.4 L_x$ \\
$\delta_\mathrm{blend}$     & $0.06 L_x$ \\
\enddata
\end{deluxetable}

We use the Newton solver with an analytic Jacobian.  We set the
tolerances as: $\epsrho = 10^{-10}$, $\epsspec = 10^{-10}$, $\epsener
= 10^{-6}$, and $\epsabs = 10^{-10}$.  We run with an
advective CFL number of $\mathcal{C} = 0.75$ and use 256 zones.
We also use the diffusion limiter described above
(Eq.~\ref{eq:difflimit}), and an additional limiter based on the
nuclear energy generation rate, which helps reduce the timestep right
as the flame is igniting.  This sets the timestep to be:
\begin{equation}
\delta t_\mathrm{nuc} = \zeta \frac{e}{\dot{S}}
\end{equation}
The idea is to not let nuclear reactions change a cell's internal
energy, $e$, by more than a fraction $\zeta$.  We use $\zeta = 0.25$ for
these simulations.  It is still possible for rapidly increasing energy
generation to violate this limiter, since we use the current
timestep's state to predict the $\delta t_\mathrm{nuc}$ for the next
step.


\begin{figure}[t]
\centering
\plotone{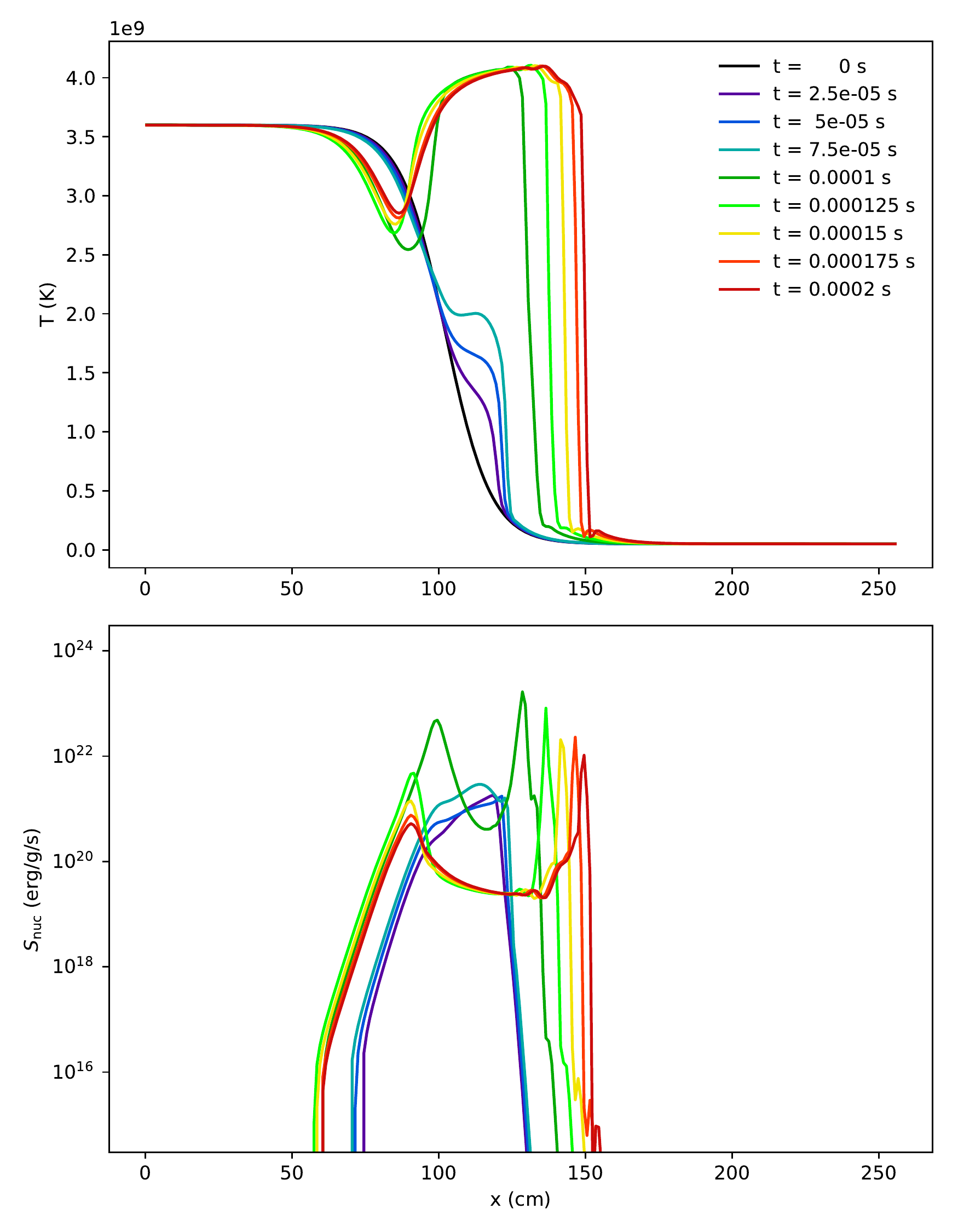}
\caption{\label{fig:flame} A helium flame run with the SDC-4 algorithm.}
\end{figure}

Figure~\ref{fig:flame} shows the temperature and energy generation
rate in flame at several instances in time, We see that diffusion and
reactions slowly increase the temperature at the fuel-ash interface
during the early evolution before the flame rapidly ignites.  At the
final point in the evolution, the flame still has not reached a steady
state.  This problem demonstrates that the SDC-4 algorithm works well with
more extensive networks.

\section{Summary}

\label{sec:summary}

We showed that spectral deferred corrections can lead to high-order
coupling of hydrodynamics and reactions for astrophysical problems.
Aside from the shock tubes and example flame, we focused on smooth
problems so we could measure convergence.  We saw that we can achieve
fourth-order convergence in reacting flow problems with stiff reaction
sources using the SDC coupling.  This work provides a path for doing
multiphysics time evolution to higher than second-order temporal
accuracy.  Extensions to this include self-gravity, including the
conservative formulation described in \citet{wdmergerI}.  We would
need to solve the Poisson equation at each time node, for each
iteration.  Using geometric multigrid, we would expect the later
iterations to converge quickly when we start with the potential from
the previous iteration.
There are also extensions to radiation, like
that explored to second-order in \citealt{sekora_higher-order_2009}.
Finally, we will expand this methodology adaptive mesh refinement with
subcycling in time.

In a follow-on paper, we will explore burning fronts more thoroughly,
including deflagrations and detonations, where it has been shown that
resolution is key to avoiding spurious numerically-seeded
detonations~\citep{katzzingale:2019}, so higher-order methods may
help.  We also want to understand how the improved coupling helps with
nuclear statistical equilibrium attained in the ashes.  Finally, our
main science target is modeling flame spreading in X-ray bursts, where
the wide range of length scales makes resolved simulations
challenging~\citep{astronum:2018}, so the push to fourth-order
reactive hydrodynamics should help.  We've demonstrated that we have
the necessary physics to fourth-order accuracy for our models of X-ray
bursts.  We focused on Cartesian geometry here.  The extension to
axisymmetric flows is straightforward, but requires deriving the
fourth-order interpolants in that geometry.  We will consider that in
a separate study.

It is straightforward to adapt an existing method-of-lines
hydrodynamics code to use this SDC integration technique.  The main
piece needed is access to the instantaneous reaction rates, instead of
relying on a network integration package.  The methodology presented
here can also be extended to radiation and implicit diffusion to
enable higher time-order and better coupling, and could be useful as
well for fully implicit hydrodynamics schemes, including those used by
stellar evolution codes.   It can also be easily adapted to cosmological
flows with chemistry.  Finally, there are a large number of
variations on the SDC approach shown here.  We could use a different
quadrature rule for the integral or subcycle on the reactions, if
needed.  We will explore variations in future papers.

\acknowledgements \castro\ is freely available at
\url{http://github.com/AMReX-Astro/Castro}.  All of the code and
problem setups used here are available in the git repo.  All runs used
\castro\ 19.08.1 \cite{castro_19_08_1}, \amrex\ 19.08
\citep{amrex_19_08}, and StarKiller Microphysics 19.08
\citep{starkiller_19_08}. A pure hydrodynamics version of the
fourth-order SDC algorithm is also implemented in the freely-available
pyro \citep{pyro,pyro_joss} code.  The work at Stony Brook was
supported by DOE/Office of Nuclear Physics grant DE-FG02-87ER40317.
This research was supported by the Exascale Computing Project
(17-SC-20-SC), a collaborative effort of the U.S. Department of Energy
Office of Science and the National Nuclear Security Administration.
The work at LBNL was supported by U.S. Department of Energy under contract No. DE-AC02-05CH11231. 
This research used resources of the National Energy Research
Scientific Computing Center (NERSC), a U.S.\ Department of Energy
Office of Science User Facility operated under Contract
No.\ DE-AC02-05CH11231.  This research has made use of NASA's
Astrophysics Data System Bibliographic Services.

\software{\amrex\ \citep{amrex_joss},
          \castro\ \citep{castro},
          GNU Compiler Collection (\url{https://gcc.gnu.org/}),
          Linux (\url{https://www.kernel.org}),
          matplotlib (\citealt{Hunter:2007},  \url{http://matplotlib.org/})
          NumPy \citep{numpy,numpy2},
          pyro \citep{pyro, pyro_joss},
          python (\url{https://www.python.org/}),
          SciPy \citep{scipy},
          StarKiller microphysics \citep{starkiller_19_08},
          SymPy \citep{sympy},
          valgrind \citep{valgrind},
          yt \citep{yt}
         }

\facility{NERSC}

\appendix

\section{Jacobian}

\label{sec:app:jac}

For solving the nonlinear update of the reacting system, we need to compute
the Jacobian, $\partial \Rb/\partial \Uc$.  We do
this in two pieces, $\partial \Rb/\partial {\bf w }$ and $\partial
{\bf w}/\partial \Uc$.  We will show two species here, called
$X_\alpha$ and $X_\beta$, so the structure of the matricies is clear
when there are multiple species.  We need to compute $\partial {\bf
  w}/\partial \Uc$, with ${\bf w} = (\rho, X_\alpha, X_\beta,
T)^\intercal$.  For this transformation, we need to pick only one of
$(\rho E)$ or $(\rho e)$.
We show the Jacobian for both choices, denoting the state as
$\Uc_{(E)}$ when we include $(\rho E)$ and as $\Uc_{(e)}$ when we include
$(\rho e)$,
\begin{equation}
\Uc_{(E)} = \left ( \begin{array}{c} \rho \\ \rho X_\alpha \\ \rho X_\beta \\ \rho E \end{array} \right )
\quad
\Uc_{(e)} = \left ( \begin{array}{c} \rho \\ \rho X_\alpha \\ \rho X_\beta \\ \rho e \end{array} \right )
\end{equation}

The Jacobian transformation $\partial \Uc/\partial {\bf w}$ for each
of these conserved state choices can be written down straightforwardly
as:
\begin{equation}
\frac{\partial \Uc_{(E)}}{\partial {\bf w}} = \left (
   \begin{array}{cccc}
       1 & 0 & 0 & 0 \\
       X_\alpha & \rho & 0 & 0\\
       X_\beta & 0 & \rho & 0 \\
       \rho e_\rho  + e + \frac{1}{2}\Ub^{2} &
                 \rho  e_{X_\alpha} & \rho e_{X_\beta} &
                 \rho e_T
     \end{array}\right)
\quad
\frac{\partial \Uc_{(e)}}{\partial {\bf w}} = \left (
   \begin{array}{cccc}
       1 & 0 & 0 & 0 \\
       X_\alpha & \rho & 0 & 0\\
       X_\beta & 0 & \rho & 0 \\
       \rho e_\rho  + e  &
                 \rho  e_{X_\alpha} & \rho e_{X_\beta} &
                 \rho e_T
     \end{array}\right)
\end{equation}
where we use the following notation for compactness:
\begin{equation}
e_\rho = \dedrd \qquad 
e_T = \dedTd \qquad
e_{X_k} = \dedXd 
\end{equation}
and the inverses (computed via SymPy, see the included Jupyter notebook) are:
\begin{equation}
\frac{\partial {\bf w}}{\partial \Uc_{(E)}} = \left (
  \begin{array}{cccc}
   1 & 0 & 0 & 0 \\
   - \frac{X_\alpha}{\rho} & \frac{1}{\rho} & 0 & 0 \\
   - \frac{X_\beta}{\rho} & 0 & \frac{1}{\rho} & 0  \\
   \left (\rho e_T \right )^{-1} \left( \sum_k X_{k} e_{X_k} - \rho e_\rho - e + \frac{1}{2}\Ub^{2}\right) &
         - \frac{e_{X_\alpha}}{\rho e_T} & - \frac{e_{X_\beta}}{\rho e_T}  & \frac{1}{\rho e_T}
   \end{array}\right)
\end{equation}
and
\begin{equation}
\frac{\partial {\bf w}}{\partial \Uc_{(e)}} = \left (
  \begin{array}{cccc}
   1 & 0 & 0 & 0 \\
   - \frac{X_\alpha}{\rho} & \frac{1}{\rho} & 0 & 0 \\
   - \frac{X_\beta}{\rho} & 0 & \frac{1}{\rho} & 0  \\
   \left (\rho e_T \right )^{-1} \left( \sum_k X_{k} e_{X_k} - \rho e_\rho - e \right ) &
         - \frac{e_{X_\alpha}}{\rho e_T} & - \frac{e_{X_\beta}}{\rho e_T}  & \frac{1}{\rho e_T}
   \end{array}\right)
\end{equation}

The reaction vector is the same regardless of the choice of $(\rho E)$ or $(\rho e)$, as
\begin{equation}
\Rb = \left (  \begin{array}{c} 0 \\ \rho \omegadot_\alpha \\ \rho \omegadot_\beta \\ \rho \Sdot \end{array} \right )
\end{equation}
and the Jacobian is computed as $\partial \Rb/\partial {\bf w}$:
\begin{equation}
\frac{\partial \Rb}{\partial {\bf w}} = \left (
  \begin{array}{cccc}
     0 & 0 & 0 & 0 \\
     \omegadot_\alpha + \rho \frac{\partial \omegadot_\alpha}{\partial \rho} &
     \rho \frac{\partial \omegadot_\alpha}{\partial X_\alpha} &
     \rho \frac{\partial \omegadot_\alpha}{\partial X_\beta} & \rho \frac{\partial \omegadot_\alpha}{\partial T} \\
     \omegadot_\beta + \rho \frac{\partial \omegadot_\beta}{\partial \rho} &
     \rho \frac{\partial \omegadot_\beta}{\partial X_\alpha} &
     \rho \frac{\partial \omegadot_\beta}{\partial X_\beta} & \rho \frac{\partial \omegadot_\beta}{\partial T} \\
     \Sdot + \rho \frac{\partial \Sdot}{\partial \rho} & \rho \frac{\partial \Sdot}{\partial X_\alpha} & \rho \frac{\partial \Sdot}{\partial X_\beta} & \rho \frac{\partial \Sdot}{\partial T} \\
  \end{array}
  \right )
\end{equation}

The first row of zeros is not as alarming as it looks, since the full Jacobian has the form
${\bf J} = {\bf I} - \delta t_m (\partial \Rb/\partial {\bf w}) (\partial {\bf w} / \partial \Uc)$.

\bibliographystyle{aasjournal}
\bibliography{ws}

\begin{thebibliography}{}
\expandafter\ifx\csname natexlab\endcsname\relax\def\natexlab#1{#1}\fi
\providecommand{\url}[1]{\href{#1}{#1}}
\providecommand{\dodoi}[1]{doi:~\href{http://doi.org/#1}{\nolinkurl{#1}}}
\providecommand{\doeprint}[1]{\href{http://ascl.net/#1}{\nolinkurl{http://ascl.net/#1}}}
\providecommand{\doarXiv}[1]{\href{https://arxiv.org/abs/#1}{\nolinkurl{https://arxiv.org/abs/#1}}}

\bibitem[{{Alastuey} \& {Jancovici}(1978)}]{alastuey:1978}
{Alastuey}, A., \& {Jancovici}, B. 1978, \apj, 226, 1034,
  \dodoi{10.1086/156681}

\bibitem[{Almgren {et~al.}(2013)Almgren, Aspden, Bell, \&
  Minion}]{Almgren:2013}
Almgren, A., Aspden, A., Bell, J., \& Minion, M. 2013, SIAM Journal on
  Scientific Computing, 35, B25, \dodoi{10.1137/110829386}

\bibitem[{{Almgren} {et~al.}(2008){Almgren}, {Bell}, {Nonaka}, \&
  {Zingale}}]{ABNZ:III}
{Almgren}, A.~S., {Bell}, J.~B., {Nonaka}, A., \& {Zingale}, M. 2008, \apj,
  684, 449, \dodoi{10.1086/590321}

\bibitem[{{Almgren} {et~al.}(2010){Almgren}, {Beckner}, {Bell}, {Day},
  {Howell}, {Joggerst}, {Lijewski}, {Nonaka}, {Singer}, \& {Zingale}}]{castro}
{Almgren}, A.~S., {Beckner}, V.~E., {Bell}, J.~B., {et~al.} 2010, \apj, 715,
  1221, \dodoi{10.1088/0004-637X/715/2/1221}

\bibitem[{Bourlioux {et~al.}(2003)Bourlioux, Layton, \& Minion}]{BLM:2003}
Bourlioux, A., Layton, A.~T., \& Minion, M.~L. 2003, Journal of Computational
  Physics, 189, 651 , \dodoi{https://doi.org/10.1016/S0021-9991(03)00251-1}

\bibitem[{Brown {et~al.}(1989)Brown, Byrne, \& Hindmarsh}]{vode}
Brown, P., Byrne, G., \& Hindmarsh, A. 1989, SIAM Journal on Scientific and
  Statistical Computing, 10, 1038, \dodoi{10.1137/0910062}

\bibitem[{{Bruenn} {et~al.}(2018){Bruenn}, {Blondin}, {Hix}, {Lentz}, {Messer},
  {Mezzacappa}, {Endeve}, {Harris}, {Marronetti}, \& {Budiardja}}]{chimera}
{Bruenn}, S.~W., {Blondin}, J.~M., {Hix}, W.~R., {et~al.} 2018, arXiv e-prints,
  arXiv:1809.05608.
\newblock \doarXiv{1809.05608}

\bibitem[{{Bryan} {et~al.}(1995){Bryan}, {Norman}, {Stone}, {Cen}, \&
  {Ostriker}}]{bryan:1995}
{Bryan}, G.~L., {Norman}, M.~L., {Stone}, J.~M., {Cen}, R., \& {Ostriker},
  J.~P. 1995, Computer Physics Communications, 89, 149,
  \dodoi{10.1016/0010-4655(94)00191-4}

\bibitem[{Byrne \& Hindmarsh(1987)}]{BYRNE19871}
Byrne, G.~D., \& Hindmarsh, A.~C. 1987, Journal of Computational Physics, 70, 1
  , \dodoi{https://doi.org/10.1016/0021-9991(87)90001-5}

\bibitem[{{Caughlan} \& {Fowler}(1988)}]{caughlan-fowler:1988}
{Caughlan}, G.~R., \& {Fowler}, W.~A. 1988, Atomic Data and Nuclear Data
  Tables, 40, 283, \dodoi{10.1016/0092-640X(88)90009-5}

\bibitem[{Colella(1985)}]{colella:1985}
Colella, P. 1985, SIAM Journal on Scientific and Statistical Computing, 6, 104,
  \dodoi{10.1137/0906009}

\bibitem[{{Colella}(1990)}]{ppmunsplit}
{Colella}, P. 1990, Journal of Computational Physics, 87, 171,
  \dodoi{10.1016/0021-9991(90)90233-Q}

\bibitem[{{Dutt} {et~al.}(2000){Dutt}, {Greengard}, \& {Rokhlin}}]{dutt:2000}
{Dutt}, A., {Greengard}, L., \& {Rokhlin}, V. 2000, BIT Numerical Mathematics,
  40, 241, \dodoi{10.1023/A:102233890}

\bibitem[{Emmett {et~al.}(2019)Emmett, Motheau, Zhang, Minion, \&
  Bell}]{Emmett:2018}
Emmett, M., Motheau, E., Zhang, W., Minion, M., \& Bell, J.~B. 2019, Combustion
  Theory and Modelling, 23, 592, \dodoi{10.1080/13647830.2019.1566574}

\bibitem[{Emmett {et~al.}(2014)Emmett, Zhang, \& Bell}]{Emmett:2014}
Emmett, M., Zhang, W., \& Bell, J.~B. 2014, Combustion Theory and Modelling,
  18, 361, \dodoi{10.1080/13647830.2014.919410}

\bibitem[{{Felker} \& {Stone}(2018)}]{athenapp}
{Felker}, K.~G., \& {Stone}, J.~M. 2018, Journal of Computational Physics, 375,
  1365, \dodoi{10.1016/j.jcp.2018.08.025}

\bibitem[{{Fryxell} {et~al.}(2000){Fryxell}, {Olson}, {Ricker}, {Timmes},
  {Zingale}, {Lamb}, {MacNeice}, {Rosner}, {Truran}, \& {Tufo}}]{flash}
{Fryxell}, B., {Olson}, K., {Ricker}, P., {et~al.} 2000, \apjs, 131, 273,
  \dodoi{10.1086/317361}

\bibitem[{{Graboske} {et~al.}(1973){Graboske}, {Dewitt}, {Grossman}, \&
  {Cooper}}]{graboske:1973}
{Graboske}, H.~C., {Dewitt}, H.~E., {Grossman}, A.~S., \& {Cooper}, M.~S. 1973,
  \apj, 181, 457

\bibitem[{{Harpole} {et~al.}(2019){Harpole}, {Zingale}, {Hawke}, \&
  {Chegini}}]{pyro_joss}
{Harpole}, A., {Zingale}, M., {Hawke}, I., \& {Chegini}, T. 2019, Journal of
  Open Source Software, 4, 1265, \dodoi{10.21105/joss.01265}

\bibitem[{Huang {et~al.}(2006)Huang, Jia, \& Minion}]{HuangJiaMinion06}
Huang, J., Jia, J., \& Minion, M. 2006, Journal of Computational Physics, 214,
  633

\bibitem[{Hunter(2007)}]{Hunter:2007}
Hunter, J.~D. 2007, Computing in Science and Engg., 9, 90,
  \dodoi{10.1109/MCSE.2007.55}

\bibitem[{{Itoh} {et~al.}(1979){Itoh}, {Totsuji}, {Ichimaru}, \&
  {Dewitt}}]{itoh:1979}
{Itoh}, N., {Totsuji}, H., {Ichimaru}, S., \& {Dewitt}, H.~E. 1979, \apj, 234,
  1079, \dodoi{10.1086/157590}

\bibitem[{Jones {et~al.}(2001--)Jones, Oliphant, Peterson, {et~al.}}]{scipy}
Jones, E., Oliphant, T., Peterson, P., {et~al.} 2001--, {SciPy}: Open source
  scientific tools for {Python}.
\newblock \url{http://www.scipy.org/}

\bibitem[{Kadioglu {et~al.}(2008)Kadioglu, Klein, \& Minion}]{KADIOGLU20082012}
Kadioglu, S.~Y., Klein, R., \& Minion, M.~L. 2008, Journal of Computational
  Physics, 227, 2012 , \dodoi{https://doi.org/10.1016/j.jcp.2007.10.008}

\bibitem[{{K{\"a}ppeli} \& {Mishra}(2016)}]{kappeli:2016}
{K{\"a}ppeli}, R., \& {Mishra}, S. 2016, \aap, 587, A94,
  \dodoi{10.1051/0004-6361/201527815}

\bibitem[{{Katz} \& {Zingale}(2019)}]{katzzingale:2019}
{Katz}, M.~P., \& {Zingale}, M. 2019, \apj, 874, 169,
  \dodoi{10.3847/1538-4357/ab0c00}

\bibitem[{{Katz} {et~al.}(2016){Katz}, {Zingale}, {Calder}, {Swesty},
  {Almgren}, \& {Zhang}}]{wdmergerI}
{Katz}, M.~P., {Zingale}, M., {Calder}, A.~C., {et~al.} 2016, \apj, 819, 94,
  \dodoi{10.3847/0004-637X/819/2/94}

\bibitem[{McCorquodale \& Colella(2011)}]{mccorquodalecolella}
McCorquodale, P., \& Colella, P. 2011, Commun. Appl. Math. Comput. Sci., 6, 1,
  \dodoi{10.2140/camcos.2011.6.1}

\bibitem[{{Meakin} \& {Arnett}(2007)}]{prompi}
{Meakin}, C.~A., \& {Arnett}, D. 2007, \apj, 667, 448, \dodoi{10.1086/520318}

\bibitem[{Meurer {et~al.}(2017)Meurer, Smith, Paprocki, \v{C}ert\'{i}k,
  Kirpichev, Rocklin, Kumar, Ivanov, Moore, Singh, Rathnayake, Vig, Granger,
  Muller, Bonazzi, Gupta, Vats, Johansson, Pedregosa, Curry, Terrel,
  Rou\v{c}ka, Saboo, Fernando, Kulal, Cimrman, \& Scopatz}]{sympy}
Meurer, A., Smith, C.~P., Paprocki, M., {et~al.} 2017, PeerJ Computer Science,
  3, e103, \dodoi{10.7717/peerj-cs.103}

\bibitem[{{Miller} \& {Colella}(2002)}]{millercolella:2002}
{Miller}, G.~H., \& {Colella}, P. 2002, Journal of Computational Physics, 183,
  26, \dodoi{10.1006/jcph.2002.7158}

\bibitem[{Minion(2003)}]{minion:2003}
Minion, M.~L. 2003, Commun. Math. Sci., 1, 471.
\newblock \url{https://projecteuclid.org:443/euclid.cms/1250880097}

\bibitem[{{Most} {et~al.}(2019){Most}, {Papenfort}, \& {Rezzolla}}]{most:2019}
{Most}, E.~R., {Papenfort}, L.~J., \& {Rezzolla}, L. 2019, arXiv e-prints,
  arXiv:1907.10328.
\newblock \doarXiv{1907.10328}

\bibitem[{{M{\"u}ller}(1986)}]{muller:1986}
{M{\"u}ller}, E. 1986, \aap, 162, 103

\bibitem[{Nethercote \& Seward(2007)}]{valgrind}
Nethercote, N., \& Seward, J. 2007, in Proceedings of the 28th ACM SIGPLAN
  Conference on Programming Language Design and Implementation, PLDI '07 (New
  York, NY, USA: ACM), 89--100, \dodoi{10.1145/1250734.1250746}

\bibitem[{{Nonaka} {et~al.}(2010){Nonaka}, {Almgren}, {Bell}, {Lijewski},
  {Malone}, \& {Zingale}}]{MAESTRO:Multilevel}
{Nonaka}, A., {Almgren}, A.~S., {Bell}, J.~B., {et~al.} 2010, \apjs, 188, 358,
  \dodoi{10.1088/0067-0049/188/2/358}

\bibitem[{Oliphant(2007)}]{numpy}
Oliphant, T.~E. 2007, Computing in Science and Engg., 9, 10,
  \dodoi{10.1109/MCSE.2007.58}

\bibitem[{Pazner {et~al.}(2016)Pazner, Nonaka, Bell, Day, \&
  Minion}]{Pazner:2016}
Pazner, W.~E., Nonaka, A., Bell, J.~B., Day, M.~S., \& Minion, M.~L. 2016,
  Combustion Theory and Modelling, 20, 521,
  \dodoi{10.1080/13647830.2016.1150519}

\bibitem[{Sekora \& Stone(2009)}]{sekora_higher-order_2009}
Sekora, M., \& Stone, J. 2009, Communications in Applied Mathematics and
  Computational Science, 4, 135, \dodoi{10.2140/camcos.2009.4.135}

\bibitem[{{Strang}(1968)}]{strang:1968}
{Strang}, G. 1968, SIAM Journal on Numerical Analysis, 5, 506,
  \dodoi{10.1137/0705041}

\bibitem[{the AMReX Development~Team {et~al.}(2019)the AMReX Development~Team,
  Almgren, Beckner, Blaschke, Chan, Day, Friesen, Gott, Graves, Katz, Myers,
  Nguyen, Nonaka, Rosso, Williams, Zhang, \& Zingale}]{amrex_19_08}
the AMReX Development~Team, Almgren, A., Beckner, V., {et~al.} 2019,
  AMReX-Codes/amrex: AMReX 19.08, \dodoi{10.5281/zenodo.3358046}

\bibitem[{the Castro Development~Team {et~al.}(2019)the Castro
  Development~Team, Almgren, Barrios~Sazo, Bell, Harpole, Katz, Willcox, Zhang,
  \& Zingale}]{castro_19_08_1}
the Castro Development~Team, Almgren, A., Barrios~Sazo, M., {et~al.} 2019,
  AMReX-Astro/Castro: Castro 19.08.1, \dodoi{10.5281/zenodo.3359184}

\bibitem[{the StarKiller Microphysics Development~Team {et~al.}(2019)the
  StarKiller Microphysics Development~Team, Bishop, Fields, Jacobs, Katz, Li,
  Malone, Timmes, Willcox, \& Zingale}]{starkiller_19_08}
the StarKiller Microphysics Development~Team, Bishop, A., Fields, C.~E.,
  {et~al.} 2019, {starkiller-astro/Microphysics: StarKiller Microphysics
  19.08}, \dodoi{10.5281/zenodo.3357970}

\bibitem[{{Timmes} \& {Swesty}(2000)}]{timmes_swesty:2000}
{Timmes}, F.~X., \& {Swesty}, F.~D. 2000, \apjs, 126, 501,
  \dodoi{10.1086/313304}

\bibitem[{{Turk} {et~al.}(2011){Turk}, {Smith}, {Oishi}, {Skory}, {Skillman},
  {Abel}, \& {Norman}}]{yt}
{Turk}, M.~J., {Smith}, B.~D., {Oishi}, J.~S., {et~al.} 2011, \apjs, 192, 9,
  \dodoi{10.1088/0067-0049/192/1/9}

\bibitem[{van~der Walt {et~al.}(2011)van~der Walt, Colbert, \&
  Varoquaux}]{numpy2}
van~der Walt, S., Colbert, S.~C., \& Varoquaux, G. 2011, Computing in Science
  \& Engineering, 13, 22, \dodoi{10.1109/MCSE.2011.37}

\bibitem[{{Wongwathanarat} {et~al.}(2016){Wongwathanarat}, {Grimm-Strele}, \&
  {M{\"u}ller}}]{apsara}
{Wongwathanarat}, A., {Grimm-Strele}, H., \& {M{\"u}ller}, E. 2016, \aap, 595,
  A41, \dodoi{10.1051/0004-6361/201628205}

\bibitem[{{Zhang} {et~al.}(2019){Zhang}, {Almgren}, {Beckner}, {Bell},
  {Blaschke}, {Chan}, {Day}, {Friesen}, {Gott}, {Graves}, {Katz}, {Myers},
  {Nguyen}, {Nonaka}, {Rosso}, {Williams}, \& {Zingale}}]{amrex_joss}
{Zhang}, W., {Almgren}, A., {Beckner}, V., {et~al.} 2019, Journal of Open
  Source Software, 4, 1370, \dodoi{10.21105/joss.01370}

\bibitem[{{Zingale}(2014)}]{pyro}
{Zingale}, M. 2014, Astronomy and Computing, 6, 52,
  \dodoi{10.1016/j.ascom.2014.07.003}

\bibitem[{{Zingale} \& {Katz}(2015)}]{zingalekatz}
{Zingale}, M., \& {Katz}, M.~P. 2015, \apjs, 216, 31,
  \dodoi{10.1088/0067-0049/216/2/31}

\bibitem[{{Zingale} {et~al.}(2002){Zingale}, {Dursi}, {ZuHone}, {Calder},
  {Fryxell}, {Plewa}, {Truran}, {Caceres}, {Olson}, {Ricker}, {Riley},
  {Rosner}, {Siegel}, {Timmes}, \& {Vladimirova}}]{ppm-hse}
{Zingale}, M., {Dursi}, L.~J., {ZuHone}, J., {et~al.} 2002, \apjs, 143, 539,
  \dodoi{10.1086/342754}

\bibitem[{Zingale {et~al.}(2019)Zingale, Eiden, Cavecchi, Harpole, Bell, Chang,
  Hawke, Katz, Malone, Nonaka, Willcox, \& Zhang}]{astronum:2018}
Zingale, M., Eiden, K., Cavecchi, Y., {et~al.} 2019, Journal of Physics:
  Conference Series, 1225, 012005, \dodoi{10.1088/1742-6596/1225/1/012005}

\end{thebibliography}

\end{document}